\documentclass[a4paper,twocolumn]{article}

\usepackage{amsmath}          
\usepackage{amstext}          
\usepackage{amssymb}          
\usepackage{amsfonts}         
\usepackage{graphicx}         
\usepackage{pict2e}           
\usepackage{color}            
\usepackage{hyperref}

\renewcommand{\O}[1]{\mathcal{O}\left(#1\right)}
\renewcommand{\P}{\operatorname{\mathbb{P}}}

\newcommand{\Id}{\mathrm{Id}}

\newcommand{\C}{\mathbb{C}}
\newcommand{\Ch}{\widehat{\C}}
\newcommand{\R}{\mathcal{R}}

\newcommand{\dd}{\mathrm{d}}
\newcommand{\ee}{\mathrm{e}}
\newcommand{\ii}{\mathrm{i}}

\let\OLDthebibliography\thebibliography
\renewcommand\thebibliography[1]{
	\OLDthebibliography{#1}
	\setlength{\parskip}{0pt}
	\setlength{\itemsep}{0pt plus 0.3ex}
}

\begin{document}
\title{Approach to stationarity for the KPZ fixed point with boundaries}
\author{Sylvain Prolhac}
\date{Laboratoire de Physique Th\'eorique, UPS, Universit\'e de Toulouse, France}


\twocolumn[
\begin{@twocolumnfalse}
	\maketitle
\begin{abstract}
	Current fluctuations for the one-dimensional totally asymmetric exclusion process (TASEP) connected to reservoirs of particles, and their large scale limit to the KPZ fixed point in finite volume, are studied using exact methods. Focusing on the maximal current phase for TASEP, corresponding to infinite boundary slopes for the KPZ height field, we obtain for general initial condition an exact expression for the late time correction to stationarity, involving extreme value statistics of Brownian paths. In the special cases of stationary and narrow wedge initial conditions, a combination of Bethe ansatz and numerical conjectures alternatively provide fully explicit exact expressions.
	
	\vspace{5mm}
\end{abstract}
\end{@twocolumnfalse}
]

\maketitle

KPZ universality in one dimension \cite{S2017.1,T2018.1} describes the large scale fluctuations of a broad class of microscopic models far from equilibrium, notably in connection with interface growth \cite{KPZ1986.1,HHZ1995.1}, classical \cite{L2023.1} and quantum \cite{PSMG2021.1,BGI2021.1,BHMKPSZ2021.1} fluids, or random geometry \cite{HH1985.1,G2022.1}. Of ongoing interest in the past few decades, the highly non-linear regime of the KPZ fixed point is characterized by universal statistics of a random field $h(x,\tau)$ at position $x$ and time $\tau\geq0$. Many exact results from the past few decades for the unimpeded growth of correlations in the idealized setting of an \emph{infinitely large system}, $x\in\mathbb{R}$, have in particular been observed with an excellent match in experiments of classical \cite{T2014.1} and quantum \cite{SSDNSGMT2021.1,WRYMKSHRGYBZ2022.1,FSBALMSLHWCARMCRB2022.1} systems.

In this letter, we consider instead the KPZ fixed point \emph{in finite volume} \cite{P2024.1}, say with $x\in[0,1]$, where the transient growth of correlations is limited by the finiteness of the system size and reaches a stationary state when $\tau\to\infty$. Since correlations eventually span the whole system size, boundary conditions matter in this setting.

With periodic boundary conditions $x\equiv x+1$, the stationary state reached at late times is a Brownian bridge $b(x)$ (i.e. a Wiener process conditioned on $b(1)=0$), and stationary large deviations are known \cite{DL1998.1}. The full time-dependent probability $\P(h(x,\tau)<u)$ has also been computed for a few simple initial states \cite{P2016.1,BL2018.1}. At short time, the Tracy-Widom distributions characteristic of the KPZ fixed point on the infinite line $x\in\mathbb{R}$ \cite{C2011.1} are recovered, while the approach to stationarity at late times involves extreme value statistics of Brownian bridges \cite{MP2018.1}.

In the more physical setting of open boundaries studied in this letter, describing e.g. driven particles flowing from one reservoir to another, see below, boundary slopes $\partial_{x}h(0,\tau)$ and $\partial_{x}h(1,\tau)$ are fixed. In this letter, we focus more specifically on \emph{infinite boundary slopes}
\begin{equation}
	\label{boundary slopes infinite}
	\partial_{x}h(0,\tau)=-\infty
	\quad\text{and}\quad
	\partial_{x}h(1,\tau)=\infty\;,
\end{equation}
which is the most natural in the context of driven particles since it does not require fine-tuning at the boundaries. The stationary state, understood as the limit $h_{\rm st}(x)=\lim_{\tau\to\infty}h(x,\tau)-h(0,\tau)$, is then equal to \cite{DEL2004.1,BLD2022.1,BWW2023.1}
\begin{equation}
	\label{hst}
	h_{\rm st}(x)=\frac{w(x)-e(x)}{\sqrt{2}}\;,
\end{equation}
where $w(x)$ is a Wiener process (Brownian motion) and $e(x)$ an independent Brownian excursion (i.e. a Wiener process $e(x)$ conditioned on $e(1)=0$ and $e(x)\geq0$ for all $x\in[0,1]$). The stationary large deviations \cite{GLMV2012.1} and the relaxation time $\tau_{*}$ to stationarity \cite{dGE2005.1,GP2020.1} are also known, but not the full dynamics of the statistics of $h(x,\tau)$. We provide a first approach to the latter problem by computing the late time relaxation to stationary,
\begin{equation}
	\label{GF[h]}
	\langle\ee^{sh(x,\tau)}\rangle\simeq\ee^{\tau\mu(s)+\nu_{x}(s)}
	\quad\text{when $\tau\to\infty$}\;.
\end{equation}
The quantity $\mu(s)$ is already known \cite{GLMV2012.1}, while the leading transient correction $\nu_{x}(s)$ is new. Successive logarithmic derivatives with respect to $s$ evaluated at $s=0$ leads in particular to the cumulants of $h(x,\tau)$,
\begin{equation}
	\label{cumulants[ck,dk]}
	\langle h(x,\tau)^{k}\rangle_{\rm c}\simeq c_{k}\,\tau+d_{k}(x)\;.
\end{equation}
The stationary cumulants $c_{k}=\mu^{(k)}(0)$ are independent of the initial condition $h(x,0)$, unlike the correction $d_{k}(x)=\nu_{x}^{(k)}(0)$, as seen on the main result (\ref{nux[h0]}) below. Additionally, we emphasize that the expansion (\ref{cumulants[ck,dk]}) of the cumulants does not have higher algebraic corrections in time: the neglected terms in the identities (\ref{GF[h]}) and (\ref{cumulants[ck,dk]}) vanish exponentially fast when $\tau$ is large compared to the relaxation time $\tau_{*}$. As such, the coefficients $d_{k}(x)$ represent the only dependency on the initial condition observable beyond a few units of $\tau_{*}\approx0.13974$ \cite{dGE2005.1,GP2020.1}.

\begin{figure}
	\setlength{\unitlength}{0.75mm}
	\begin{picture}(100,57)(-8,-4)
		\put(0,0){\line(1,0){100}}
		\multiput(0,0)(10,0){11}{\line(0,1){5}}
		\put(-1,-4){$0$}
		\put(9,-4){$1$}
		\put(19,-4){$2$}
		\put(98.5,-4){$L$}
		\put(15,3){\circle*{2}}
		\put(25,3){\circle*{2}}
		\put(35,3){\circle*{2}}
		\put(65,3){\circle*{2}}
		\put(85,3){\circle*{2}}
		\put(95,3){\circle*{2}}
		\qbezier(-5,5)(0,10)(5,5)
		\put(5,5){\vector(1,-1){0.2}}
		\put(-1,8.5){$\alpha$}
		\qbezier(35,5)(40,10)(45,5)
		\put(45,5){\vector(1,-1){0.2}}
		\put(39,8.5){$1$}
		\qbezier(65,5)(70,10)(75,5)
		\put(75,5){\vector(1,-1){0.2}}
		\put(69,8.5){$1$}
		\qbezier(95,5)(100,10)(105,5)
		\put(105,5){\vector(1,-1){0.2}}
		\put(99,8.5){$\beta$}
		\multiput(0,13.75)(0,2.5){9}{\color[rgb]{0.7,0.7,0.7}\!.}
		\multiput(10,6.25)(0,2.5){8}{\color[rgb]{0.7,0.7,0.7}\!.}
		\multiput(20,6.25)(0,2.5){4}{\color[rgb]{0.7,0.7,0.7}\!.}
		\multiput(30,6.25)(0,2.5){8}{\color[rgb]{0.7,0.7,0.7}\!.}
		\multiput(40,13.75)(0,2.5){1}{\color[rgb]{0.7,0.7,0.7}\!.}
		\multiput(50,6.25)(0,2.5){8}{\color[rgb]{0.7,0.7,0.7}\!.}
		\multiput(60,6.25)(0,2.5){4}{\color[rgb]{0.7,0.7,0.7}\!.}
		\multiput(70,13.75)(0,2.5){5}{\color[rgb]{0.7,0.7,0.7}\!.}
		\multiput(80,6.25)(0,2.5){4}{\color[rgb]{0.7,0.7,0.7}\!.}
		\multiput(90,6.25)(0,2.5){8}{\color[rgb]{0.7,0.7,0.7}\!.}
		\multiput(100,13.75)(0,2.5){1}{\color[rgb]{0.7,0.7,0.7}\!.}
		\put(0,15){\color[rgb]{0.7,0.7,0.7}\line(1,1){20}}
		\put(20,15){\color[rgb]{0.7,0.7,0.7}\line(1,1){10}}
		\put(60,15){\color[rgb]{0.7,0.7,0.7}\line(1,1){10}}
		\put(80,15){\color[rgb]{0.7,0.7,0.7}\line(1,1){10}}
		\put(20,15){\color[rgb]{0.7,0.7,0.7}\line(-1,1){20}}
		\put(60,15){\color[rgb]{0.7,0.7,0.7}\line(-1,1){10}}
		\put(80,15){\color[rgb]{0.7,0.7,0.7}\line(-1,1){10}}
		\put(0,35){\thicklines\line(1,1){10}}
		\put(10,45){\thicklines\line(1,-1){30}}
		\put(40,15){\thicklines\line(1,1){20}}
		\put(60,35){\thicklines\line(1,-1){10}}
		\put(70,25){\thicklines\line(1,1){10}}
		\put(80,35){\thicklines\line(1,-1){20}}
		\put(0,47){\vector(0,-1){9}}
		\put(1.5,42.5){$\alpha$}
		\put(30,40){\line(1,1){10}}
		\put(30,40){\line(1,-1){10}}
		\put(40,50){\line(1,-1){10}}
		\put(40,30){\line(1,1){10}}
		\put(40,27){\vector(0,-1){9}}
		\put(41.5,21.5){$1$}
		\put(70,37){\vector(0,-1){9}}
		\put(71.5,31.5){$1$}
		\put(90,40){\line(1,1){10}}
		\put(90,40){\line(1,-1){10}}
		\put(100,30){\color[rgb]{0.7,0.7,0.7}\line(0,1){20}}
		\put(100,27){\vector(0,-1){9}}
		\put(101.5,21.5){$\beta$}
	\end{picture}
	\caption{TASEP with open boundaries (bottom), and the corresponding discrete height function $H_{i}(t)$ (top).}
	\label{fig TASEP}
\end{figure}
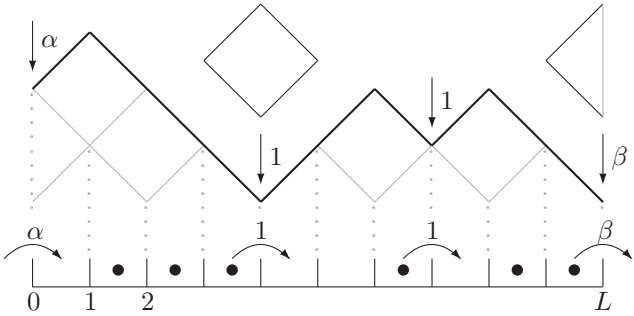

The computations in this letter are performed on a well known integrable discretization of the KPZ fixed point called the totally asymmetric simple exclusion process (TASEP), before taking the KPZ scaling limit. TASEP is a Markov process where hard-core particles hop from one site $i$ to the next site $i+1$ (if the latter is empty) on a one-dimensional lattice of $L$ sites, see figure~\ref{fig TASEP}. The hopping rate of particles in the bulk of the system is set to $1$. Additionally, particles can enter (resp. leave) the lattice at site $i=1$ (resp. from site $i=L$) with rate $\alpha$ (resp. $\beta$) if that site is empty (resp. occupied).

A classical mapping of TASEP to an interface growth model, see figure~\ref{fig TASEP}, interprets empty (resp. occupied) sites as portions of interface going up (resp. down). Modulo a global shift of the interface, this defines a bijection between configurations of the particles on the lattice and discrete heights functions $H_{i}$, normalized with $H_{0}=0$, written from occupation numbers $n_{i}$ (with $n_{i}=1$ if site $i$ is occupied and $n_{i}=0$ if site $i$ is empty) as
\begin{equation}
	\label{H[C]}
	H_{i}=\sum_{\ell=1}^{i}\Big(\frac{1}{2}-n_{\ell}\Big)\;,
\end{equation}
$i=0,\ldots,L$. Movements of particles on the lattice then induce deposition of square blocks on the local minima of the interface, with $H_{i}\to H_{i}+1$ when a particle moves from site $i$ to site $i+1$ (with $i=0$ for particles entering at site $1$ and $i=L$ for particles leaving from site $L$). The resulting time dependent height function $H_{i}(t)$ is initialized at time $t=0$ as (\ref{H[C]}) with the occupation numbers of the initial configuration (which may be random). Additionally, the time-integrated current $Q_{i}(t)$, defined as the total number of particles that have moved from site $i$ to $i+1$ up to time $t$, is equal to
\begin{equation}
	\label{Q[H]}
	Q_{i}(t)=H_{i}(t)-H_{i}(0)\;.
\end{equation}
Incidentally, the identity $H_{i}(t)=Q_{0}(t)+\sum_{\ell=1}^{i}\big(\frac{1}{2}-n_{\ell}(t)\big)$ holds, and there is thus a proper bijection at any time $t$ between height functions $H_{i}(t)$ and the joint data of the current at a given site and the configuration of particles.

In the whole maximal current phase $\alpha>1/2$, $\beta>1/2$ of TASEP \cite{K1991.1,DEHP1993.1,SD1993.1}, the large scale limit $L\to\infty$ of the discrete height on the time scale $t\sim L^{3/2}$ leads to the KPZ fixed point with infinite boundary slopes, and we set already $\alpha=\beta=1$ for simplicity in the following. Considering a smooth enough continuous function $h_{0}(x)$, $0\leq x\leq1$, with $h_{0}(0)=0$, we choose an initial condition of the form
\begin{equation}
	\label{H0[h0]}
	H_{i=xL}(0)\simeq\frac{\sqrt{L}}{2}\,h_{0}(x)\;,
\end{equation}
Then, on the time scale
\begin{equation}
	\label{t[tau]}
	t=2\tau L^{3/2}\;,
\end{equation}
the KPZ height $h(x,\tau)$ appears after subtracting the deterministic stationary growth $t/4$ as
\begin{equation}
	\label{H[h]}
	H_{i=xL}(t)\underset{L\to\infty}{\simeq}\frac{t}{4}+\frac{\sqrt{L}}{2}\,h(x,\tau)\;.
\end{equation}

The statistics of the TASEP current $Q_{i}(t)$ is conveniently studied from the generating function \cite{DL1998.1}
\begin{equation}
	\label{GF[M]}
	\langle \ee^{\gamma\,Q_{i}(t)}\rangle=\langle\Sigma|\ee^{tM_{i}(\gamma)}|\mathbb{P}_{0}\rangle\;.
\end{equation}
The matrix elements of the generator $M_{i}(\gamma)$ are equal to those of the Markov operator of TASEP, multiplied by $\ee^{\gamma}$ for all transitions where a particle moves from site $i$ to site $i+1$. One has in particular $M_{i}(\gamma)=V_{i}^{-1}(\gamma)M_{0}(\gamma)V_{i}(\gamma)$ with $V_{i}$ diagonal such that $\langle C|V_{i}|C\rangle=\ee^{-\gamma H_{i}}$, and $H_{i}$ related to the configuration $C$ of particles on the lattice through (\ref{H[C]}). The row vector $\langle\Sigma|=(1,1,\ldots,1)$ in (\ref{GF[M]}) leads to a summation over all possible final states, and $|\mathbb{P}_{0}\rangle$ is the column vector of initial probabilities.

Expanding (\ref{GF[M]}) over the eigenstates of $M_{i}(\gamma)$, only the contribution of what we refer to by abuse of language as the \emph{stationary eigenstate} of $M_{i}(\gamma)$ is needed at late times. For $\gamma\in\mathbb{R}$, the relevant eigenvalue $E_{\rm st}(\gamma)$ is the one with largest real part. Calling $\langle\psi_{\rm st}^{i}(\gamma)|$ and $|\psi_{\rm st}^{i}(\gamma)\rangle$ the corresponding left and right eigenvectors, one has
\begin{equation}
	\label{GF[psi]}
	\langle\ee^{\gamma\,Q_{i}(t)}\rangle\underset{t\to\infty}{\simeq}\frac{\langle\Sigma|\psi_{\rm st}^{i}(\gamma)\rangle\langle\psi_{\rm st}^{i}(\gamma)|\mathbb{P}_{0}\rangle}{\langle\psi_{\rm st}^{i}(\gamma)|\psi_{\rm st}^{i}(\gamma)\rangle}\,\ee^{tE_{\rm st}(\gamma)}\;.
\end{equation}

We use in the following two complementary approaches in order to extract from (\ref{GF[psi]}) the function $\nu_{x}(s)$ in (\ref{GF[h]}). First, an iterative matrix product representation \cite{L2013.1} for $\langle\psi_{\rm st}^{0}(\gamma)|$ and $|\psi_{\rm st}^{0}(\gamma)\rangle$, which leads, following \cite{MP2018.1}, to a representation of $\nu_{x}(s)$ in terms of Brownian paths valid for arbitrary initial condition $h_{0}(x)$. Then, for $x=0$, the Bethe equations conjectured in \cite{CN2018.1} for $E_{\rm st}(\gamma)$, from which the contribution of the stationary eigenstate in (\ref{GF[psi]}) is guessed for simple initial conditions by analogy with earlier results \cite{P2020.2} for periodic boundaries.

We begin with the iterative, perturbative construction in \cite{L2013.1} of the stationary eigenvector of $M_{0}(\gamma)$. Up to arbitrary order $n\in\mathbb{N}$ in $\gamma$, one has
\begin{align}
	\label{psi[T,U]}
	|\psi_{\rm st}^{0}(\gamma)\rangle&=\big(U\,T(\gamma)\big)^{n}\,U|\Sigma\rangle+\mathcal{O}(\gamma^{n+1})\nonumber\\
	\langle\psi_{\rm st}^{0}(\gamma)|&=\langle\Sigma|\big(U\,T(\gamma)\big)^{n}+\mathcal{O}(\gamma^{n+1})\;.
\end{align}
The entries of the row and column vectors $\langle\Sigma|$ and $|\Sigma\rangle$ are equal to $1$, while those of the matrices $T(\gamma)$ and $U$ have a matrix product representation \cite{L2013.1}. Following \cite{MP2018.1}, we find the alternative height representation
\begin{equation}
	\label{T[H]}
	\langle C^{2}|T(\gamma)|C^{1}\rangle=\ee^{-\gamma\,\max(H^{2}-H^{1})}\;,
\end{equation}
where height functions $H_{i}^{1}$, $H_{i}^{2}$ are respectively associated with configurations $C^{1}$, $C^{2}$ under the mapping (\ref{H[C]}), and $\max(H^{2}-H^{1})\geq0$ is the maximum of $H_{i}^{2}-H_{i}^{1}$ over all $i=0,\ldots,L$. Similarly, $\langle C^{2}|U|C^{1}\rangle$ is equal to $1$ if $H_{L}^{1}=H_{L}^{2}$ and $H_{i}^{1}\geq H_{i}^{2}$ for all $i$, and $0$ otherwise.

Inserting intermediate configurations in the product of matrices $T(\gamma)$, $U$ in (\ref{psi[T,U]}) gives a summation over discrete height functions, interpreted at large $L$ by Donsker's theorem, see e.g. \cite{D1999.1}, as an averaging over independent Wiener processes $w(x)$ as
\begin{equation}
	\label{Donsker}
	H_{i=xL}\simeq\frac{\sqrt{L}}{2}\,w(x)\;.
\end{equation}

On the KPZ time scale (\ref{t[tau]}), the eigenvalues of $M_{i}(\gamma)$ are well separated on the scale $L^{-3/2}$ \cite{GP2020.1}, so that (\ref{GF[h]}) can be extracted from the contribution (\ref{GF[psi]}) of the stationary eigenstate of TASEP. The corresponding eigenvalue
\begin{equation}
	\label{E[mu] ; gamma[s]}
	E_{\rm st}(\gamma)\simeq\frac{\gamma}{4}+\frac{\mu(s)}{2L^{3/2}}
	\quad\text{with}\quad
	\gamma=\frac{2s}{\sqrt{L}}
\end{equation}
contributes the leading term $\mu(s)$ in (\ref{GF[h]}), with the known parametric expression \cite{GLMV2012.1} $\mu(s)=s+\chi(v)/2$, $s=\chi'(v)$,
\begin{equation}
	\label{chi}
	\chi(v)=-\frac{1}{4\sqrt{\pi}}\sum_{k=1}^{\infty}\frac{(2k)!}{k^{k+5/2}\,k!}\,\Big(-\frac{\ee^{v}}{4}\Big)^{k}\;.
\end{equation}

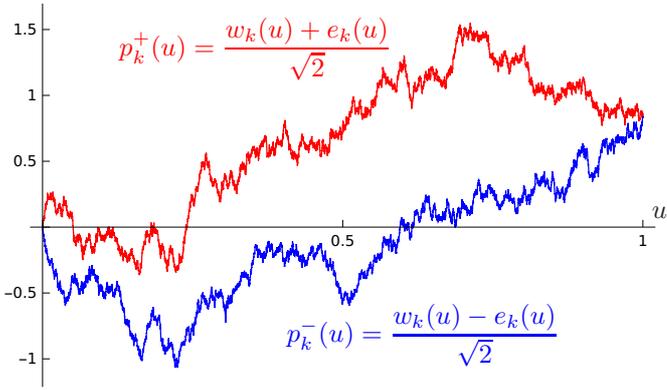
\begin{figure}
	\includegraphics[width=86mm]{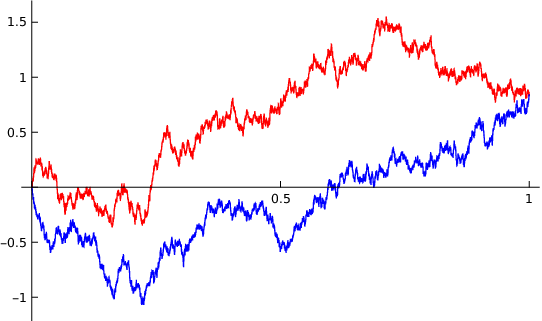}
	\begin{picture}(0,0)
		\put(-2,22.5){$u$}
		\put(-72,44){\color{red}$\displaystyle p_{k}^{+}(u)=\frac{w_{k}(u)+e_{k}(u)}{\sqrt{2}}$}
		\put(-50,6){\color{blue}$\displaystyle p_{k}^{-}(u)=\frac{w_{k}(u)-e_{k}(u)}{\sqrt{2}}$}
	\end{picture}
	\vspace{-3.5mm}\\
	\caption{Typical realization of the random paths $p_{k}^{\pm}$ in the main result (\ref{nux[h0]}), such that $p_{k}^{\pm}(0)=0$ and $p_{k}^{+}(0)=p_{k}^{-}(0)$.}
	\label{fig Brownian paths}
\end{figure}

For the correction $\nu_{x}(s)$ in (\ref{GF[h]}), Donsker's theorem (\ref{Donsker}) leads, with the scaling in (\ref{E[mu] ; gamma[s]}) for $\gamma$, to a representation of the matrix $T(\gamma)$ in (\ref{psi[T,U]}) in terms of Wiener processes,
\begin{equation}
	\label{T[w]}
	\langle C^{2}|T(\gamma)|C^{1}\rangle\simeq\ee^{s\min(w_{1}-w_{2})}\;,
\end{equation}
where $w_{1}(u)$ and $w_{2}(u)$ are associated with $C^{1}$ and $C^{2}$, and the minimum is taken over all $u\in[0,1]$. Similarly, $\langle C^{2}|U|C^{1}\rangle$ leads to conditioning the corresponding $w_{1}$ and $w_{2}$ on $w_{1}(u)\geq w_{2}(u)$ for $u\in[0,1]$ and $w_{1}(1)=w_{2}(1)$. Equivalently, $w(x)=(w_{1}(x)+w_{2}(x))/\sqrt{2}$ and $e(x)=(w_{1}(x)-w_{2}(x))/\sqrt{2}$ are respectively a Wiener process and a Brownian excursion, independent from each other, see the Supplementary Material \cite{SM}. Then, (\ref{GF[psi]}) combined with (\ref{Q[H]}), (\ref{psi[T,U]}), (\ref{T[H]}) and (\ref{Donsker}) finally leads to our first main result, valid for any $n\in\mathbb{N}$ and arbitrary initial condition $h_{0}$
\begin{equation}
	\label{nux[h0]}
	\ee^{\nu_{x}(s)}=\frac{\Big\langle\ee^{s\min(h_{0}-p_{1}^{+})+s\sum\limits_{k=1}^{n-1}\min(p_{k}^{-}-p_{k+1}^{+})}\Big\rangle}{Z_{2n}(s;0)/Z_{n}(s;x)}+\O{s^{n+1}}
\end{equation}
with $Z_{m}(s;x)=\langle\ee^{s\sum_{k=1}^{m}\min(p_{k}^{-}-p_{k+1}^{+})+s\,p_{n+1}^{-}(x)}\rangle$. The random paths $p_{k}^{\pm}(u)=\frac{w_{k}(u)\pm e_{k}(u)}{\sqrt{2}}$, such that individually $\mp p_{k}^{\pm}$ has the same statistics as $h_{\rm st}$, are defined from independent Wiener processes $w_{k}(u)$ and Brownian excursions $e_{k}(u)$, $k\in\mathbb{N}$, see figure~\ref{fig Brownian paths}. The minima in (\ref{nux[h0]}) are taken over positions $u\in[0,1]$. Following \cite{MP2018.1}, we note that (\ref{nux[h0]}) can also be reformulated in terms of conditional probabilities of non-intersecting paths $p_{k}^{\pm}$, shifted by exponentially distributed random variables with parameter $s$, which is reminiscent of line ensembles appearing for KPZ fluctuations on the infinite line $x\in\mathbb{R}$ \cite{S2006.1}.

As expected physically, the dependency in the position $x$ of the cumulants (\ref{cumulants[ck,dk]}) decouples from the initial condition at late times. Note however that the temporal and spatial increments of the height $h(0,\tau)$ and $h(x,\tau)-h(0,\tau)$ remain coupled even at late times.

\begin{figure}[t]
	\includegraphics[width=86mm]{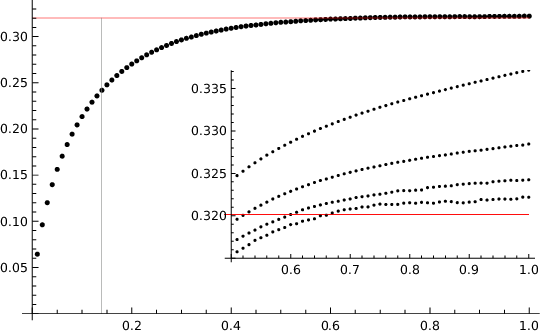}
	\begin{picture}(0,0)
		\put(-2,4){$\tau$}
		\put(-72,0){$\tau_{*}$}
		\put(-44,44){$\mathrm{Var}(h(0,\tau))-c_{2}\,\tau$\quad stat}
	\end{picture}\\
	\includegraphics[width=86mm]{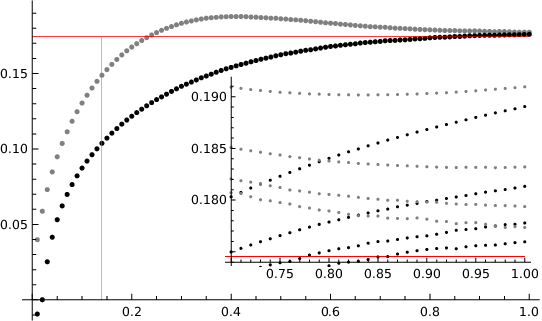}
	\begin{picture}(0,0)
		\put(-2,4.5){$\tau$}
		\put(-72,0.5){$\tau_{*}$}
		\put(-39,39){$\mathrm{Var}(h(0,\tau))-c_{2}\,\tau$\quad nw}
	\end{picture}
	\vspace{-3.5mm}\\
	\caption{$\mathrm{Var}(h(0,\tau))-c_{2}\,\tau$ with $c_{2}=\frac{3\sqrt{\pi}}{4\sqrt{2}}$ plotted as a function of time for stationary (top) and narrow wedge (bottom) initial condition. The data were obtained from simulations of open TASEP with $L=1023$ sites, boundary rates $\alpha=\beta=1$, averaged over $2^{25}$ realizations, for stationary (top), empty (bottom, gray dots) and full (bottom, black dots) initial condition. The red, horizontal lines are the exact expressions (\ref{d12(0) stat}) and (\ref{d12(0) nw}) for $d_{2}(0)$. As seen in the inset, which represents from top to bottom simulations with $L=127,255,511,1023$, the small discrepancy observed at late times appears to be consistent with a finite size correction $\sim1/L$.}
	\label{fig simul var TASEP}
\end{figure}
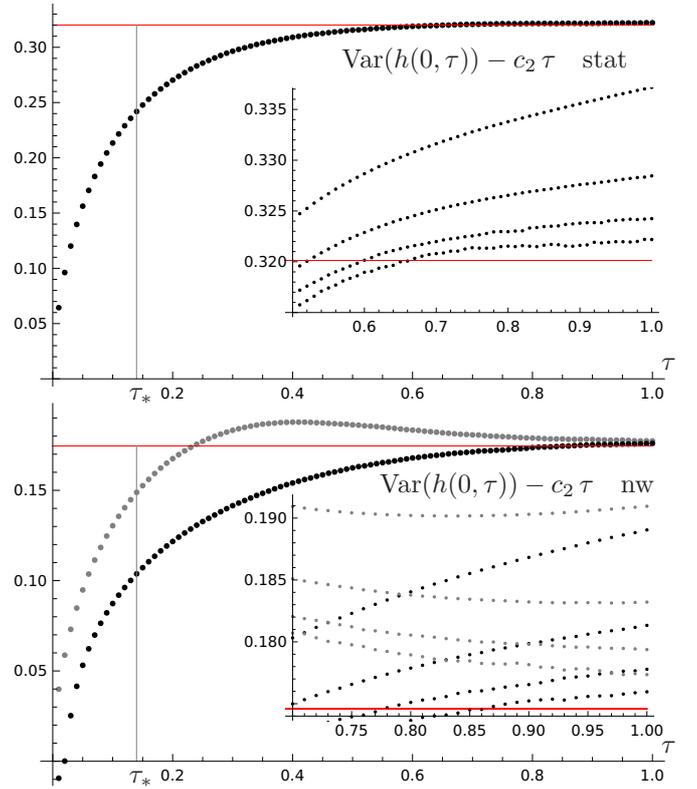

The average height $\langle h(x,\tau)\rangle$ has the form (\ref{cumulants[ck,dk]}) with the known value $c_{1}=3/2$. From (\ref{nux[h0]}), the new correction to stationarity (see the Supplementary Material \cite{SM} for the computation of the required Brownian averages) is
\begin{equation}
	\label{d1[h0]}
	d_{1}(x)=\frac{3\sqrt{\pi}}{2\sqrt{2}}-\frac{2\sqrt{x(1-x)}}{\sqrt{\pi}}+\Big\langle\min\Big(h_{0}+h_{\rm st}\Big)\Big\rangle\;,
\end{equation}
with $h_{\rm st}$ given in (\ref{hst}) and independent from $h_{0}$ when the latter is random. Fully explicit values are obtained for special initial conditions. In the case of stationary initial condition $h_{0}=h_{\rm st}$, one finds
\begin{equation}
	\label{d12(0) stat}
	d_{1}^{\rm stat}(0)=0
	\quad\text{and}\quad
	d_{2}^{\rm stat}(0)=4+\frac{9\pi}{4}-\frac{160\pi}{27\sqrt{3}}\;.
\end{equation}
For a (half) narrow wedge at the left boundary, corresponding to $h_{0}(u)\to\infty$ for $u\in(0,1]$ and thus $\min(h_{0}-p_{1}^{+})=0$ in (\ref{nux[h0]}), one finds instead
\begin{equation}
	\label{d12(0) nw}
	d_{1}^{\rm nw}(0)=\frac{3\sqrt{\pi}}{2\sqrt{2}}
	\quad\text{and}\quad
	d_{2}^{\rm nw}(0)=4+\frac{45\pi}{8}-\frac{320\pi}{27\sqrt{3}}\;.
\end{equation}
A good agreement is found with simulations of TASEP, up to finite size corrections, see figure~\ref{fig simul var TASEP}.

We now move on to the second, Bethe ansatz approach considered in this letter for the computation of the correction to stationarity in (\ref{GF[h]}) at $x=0$. Bethe ansatz postulates exact expressions for the eigenvectors of the generator of special, integrable model (such as the Hamiltonian of some quantum spin chains, or $M_{i}(\gamma)$ for TASEP). The arbitrary momenta appearing in this ansatz generally satisfy algebraic equations, called the Bethe equations.

In the case of open TASEP considered here, the Bethe equations were guessed in \cite{CN2018.1}, and the actual ``ansatz'' part for the eigenvectors is still missing, which prevents in principle the evaluation of (\ref{GF[psi]}). For $i=0$ and boundary rates $\alpha=\beta=1$, however, we have managed to guess the exact contribution of the eigenvectors in (\ref{GF[psi]}) for simple initial conditions by analogy with known corresponding expressions for periodic boundaries. These conjectures were checked numerically with high precision.

According to \cite{CN2018.1}, the stationary eigenstate of $M_{i}(\gamma)$ has the eigenvalue $E_{\rm st}(\gamma)=\frac{1}{2}\sum_{j=0}^{L+1}\frac{y_{j}}{1-y_{j}}$, where the Bethe roots $y_{j}$, which are distinct, form a solution of the Bethe equations. With the variables $w_{j}=(y_{j}+y_{j}^{-1})/2$, those Bethe equations read for $\alpha=\beta=1$
\begin{equation}
	\label{Bethe eq w}
	2^{L+2}B(1-w_{j})^{L+1}(1+w_{j})=1\;.
\end{equation}
For a given value of $\gamma$, the parameter $B$ is determined self-consistently from $\ee^{\gamma}=\prod_{j=0}^{L+1}\frac{1}{1-y_{j}}$. A comparable structure is known for periodic boundaries, involving a similar parameter $B$, see e.g. \cite{P2020.2}.

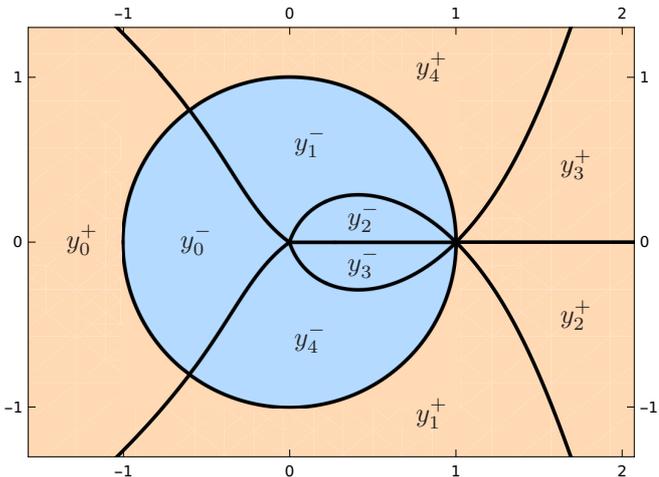
\begin{figure}
	\includegraphics[width=86mm]{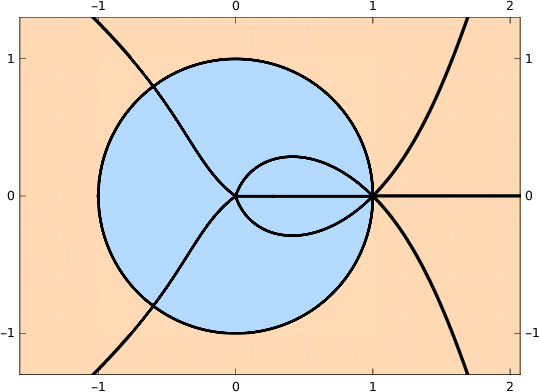}
	\begin{picture}(0,0)
		\put(-79,30.5){$y_{0}^{+}$}
		\put(-64,30.5){$y_{0}^{-}$}
		\put(-33,7.5){$y_{1}^{+}$}
		\put(-49,43.5){$y_{1}^{-}$}
		\put(-14,20.5){$y_{2}^{+}$}
		\put(-42,33.5){$y_{2}^{-}$}
		\put(-14,40.5){$y_{3}^{+}$}
		\put(-42,27.5){$y_{3}^{-}$}
		\put(-33,53.5){$y_{4}^{+}$}
		\put(-49,17.5){$y_{4}^{-}$}
	\end{picture}
	\vspace{-3.5mm}\\
	\caption{Image of the slit plane $\C\setminus[0,\infty)$ by the Bethe root functions $y_{j}^{\pm}(B)$ for open TASEP with $L=3$ and $\alpha=\beta=1$.}
	\label{fig yj}
\end{figure}

The $L+2$ solutions $w_{j}$, $j=0,\ldots,L+1$ of (\ref{Bethe eq w}) lead to $y_{j}=y_{j}^{\pm}=w_{j}\pm w_{j}\sqrt{1-w_{j}^{-2}}$. Taking the principal branch for the square root, we note that the image of the branch cut $w_{j}\in(-\infty,-1]\cup[1,\infty)$ is the unit circle $|y_{j}^{\pm}|=1$, and one has $|y_{j}^{+}|>1$, $|y_{j}^{-}|<1$ outside the cut, see figure~\ref{fig yj}. Our numerics indicate that any eigenstate of $M_{i}(\gamma)$ can be generated by choosing Bethe roots $y_{j}=y_{j}^{\theta_{j}}$ with the signs $\theta_{j}=\pm$ such that
\begin{equation}
	\label{condition thetaj}
	\prod_{j=0}^{L+1}\theta_{j}=(-1)^{L}\;.
\end{equation}
The stationary eigenstate corresponds in particular to selecting $y_{j}=y_{j}^{-}$ for all $j$.

As a function of the parameter $B$, the Bethe roots $y_{j}^{\pm}$ are analytic outside the interval $[0,\infty)$, with the three branch points $0$, $B_{*}>0$, $\infty$, where
\begin{equation}
	\label{B*}
	B_{*}=\frac{(L+2)^{L+2}}{2^{2L+4}(L+1)^{L+1}}\;.
\end{equation}
Analytic continuation across the two branch cuts $(0,B_{*})$ and $(B_{*},\infty)$ connects the various eigenstates of $M_{i}(\gamma)$ on the spectral curve $\det(\lambda\,\Id-M_{i}(\gamma))=0$. Labelling the $y_{j}^{\pm}$ as in figure~\ref{fig yj}, we observe that analytic continuation across the cuts $(B_{*},\infty)$ and $(0,B_{*})$ is realized by the action on the set $\{j_{\theta_{j}},j=0,\ldots,L+1\}$ of the respective permutations (written as a product of two cycles) $\mathfrak{a}=(0_{+},0_{-})\,(1_{+},2_{+},\ldots,(L+1)_{+},1_{-},2_{-},\ldots,(L+1)_{-})$ and $\mathfrak{b}=(0_{+},1_{+},\ldots,(L+1)_{+})\,(0_{-},1_{-},\ldots,(L+1)_{-})$. The condition (\ref{condition thetaj}) for acceptable choices of the Bethe roots $y_{j}=y_{j}^{\theta_{j}}$ is in particular preserved by $\mathfrak{a}$ and $\mathfrak{b}$, which gives an independent check of its validity.

For a given choice of signs $\theta_{j}$, writing $\langle\psi|$ and $|\psi\rangle$ for the corresponding left and right eigenvectors of $M_{0}(\gamma)$, we consider the function $\mathcal{N}(B)=\frac{\langle\Sigma|\psi\rangle\langle\psi|\mathbb{P}_{0}\rangle}{\langle\psi|\psi\rangle}$. It was shown in \cite{P2020.2} for TASEP with periodic boundaries that the corresponding meromorphic differential $\omega=\dd\log(\kappa\,\mathcal{N})$ with $\kappa=\frac{\dd\gamma}{\dd\log B}$ has a simple expression in terms of $\kappa$ and $\gamma$ for a few simple initial conditions. Trying analogous expressions for open TASEP with $\alpha=\beta=1$, we obtained the following conjectures, respectively for stationary initial condition and for the deterministic initial conditions with a system either totally empty or filled with particles,
\begin{align}
	\label{omega stat}
	\omega^{\rm stat}&=\Big((L+2)\,\kappa^{2}+\frac{\kappa}{\tanh(\gamma/2)}-1\Big)\frac{\dd B}{B}\\
	\label{omega empty}
	\omega^{\rm empty}&=\Big((L+2)\,\kappa^{2}-\frac{\kappa}{1-\ee^{\gamma}}-\kappa\Big)\frac{\dd B}{B}
\end{align}
and $\omega^{\rm full}=\omega^{\rm empty}-\frac{L}{2}\,\kappa\,\frac{\dd B}{B}$. These conjectures were checked against numerical diagonalization of $M_{0}(\gamma)$ up to system size $L=7$, with $100$ digits of precision, for a few generic values of $B$, and for all acceptable choices of signs $\theta_{j}$ satisfying (\ref{condition thetaj}).

When $L\to\infty$ under KPZ scaling (\ref{E[mu] ; gamma[s]}) for $\gamma$, setting $B/B_{*}=-\ee^{v-1}$, with $B_{*}$ in (\ref{B*}), a residue computation leveraging the fact that only the $y_{j}^{-}$ vanish when $B\to0$ leads for the stationary eigenstate to $\kappa_{\rm st}\simeq2\chi''(v)/\sqrt{L}$ with $\chi$ as in (\ref{chi}) and $v<1$ the unique solution of $\chi'(v)=s$ for $s\in(0,s_{*}]$ with $s_{*}=\chi'(1)\approx0.145931$. Noting that $\mathcal{N}(B)$ evaluated at the stationary eigenstate appears in (\ref{GF[psi]}), we finally obtain our second main result for infinite boundary slopes,
\begin{equation}
	\label{nu0 nw}
	\ee^{\nu_{0}^{\rm nw}(s)}=\frac{\chi'(v)}{\chi''(v)}\,\exp\Big(4\int_{-\infty}^{v}\dd v\,\chi''(v)^{2}\Big)
\end{equation}
for narrow wedge initial condition at the left boundary, using $\omega^{\rm empty}$ in (\ref{omega empty}). The same result is obtained for a narrow wedge at the right boundary, using $\omega^{\rm full}$, with an additional deterministic subtraction in (\ref{H[h]}), see the Supplementary Material \cite{SM}. For stationary initial condition $h_{0}=h_{\rm st}$ defined in (\ref{hst}), one finds instead from (\ref{omega stat})
\begin{equation}
	\label{nu0 stat}
	\ee^{\nu_{0}^{\rm stat}(s)}=8\sqrt{\pi}\,\ee^{-v}\chi'(v)\,\ee^{\nu_{0}^{\rm nw}(s)}\;.
\end{equation}
Expanding near $s=0$, corresponding to $v\to-\infty$, we recover the exact corrections in (\ref{d12(0) stat}) and (\ref{d12(0) nw}). Explicit expressions for the higher cumulants of $h(0,\tau)$ are also extracted easily from (\ref{nu0 nw}), (\ref{nu0 stat}), without the need to perform the complicated Gaussian integrals with non-trivial boundaries resulting from the Brownian averages in (\ref{nux[h0]}).

Of independent interest, we note that the identity between (\ref{nux[h0]}) with $h_{0}\to\infty$ and $h_{0}=h_{\rm st}$ and (\ref{nu0 nw}), (\ref{nu0 stat}) provides new conjectural formulas for generating functions of extreme value statistics of Brownian paths, complementing the ones already obtained in \cite{MP2018.1} for Brownian bridges from KPZ fluctuations with periodic boundaries.

\noindent\textit{Conclusions}. -- The exact formulas reported in this letter for the late time correction to stationarity in (\ref{GF[h]}), (\ref{cumulants[ck,dk]}), first as the Brownian average (\ref{nux[h0]}) for general initial condition, then as the explicit formulas (\ref{nu0 nw}), (\ref{nu0 stat}) for narrow wedge and stationary initial conditions, are the first results describing the universal transient dynamics of KPZ fluctuations in finite volume with open boundaries. Analytic continuation to higher eigenstates, which connects relaxation times with stationary large deviations $\mu(s)$ \cite{GP2020.1}, appears promising for the description of the complete transient dynamics of the KPZ height field $h(x,\tau)$.


\newpage
\onecolumn
\begin{center}
	\textbf{\large Supplementary Material}
\end{center}
\setcounter{page}{1}
\makeatletter
\renewcommand{\theequation}{S\arabic{equation}}
\renewcommand{\thefigure}{S\arabic{figure}}
\setcounter{secnumdepth}{3}
\def\thesection{\Roman{section}}
\def\thesubsection{\Roman{section}.\arabic{subsection}}
\def\thesubsubsection{\Roman{section}.\arabic{subsection}.\alph{subsubsection}}

We give in this supplementary material a more detailed derivation of the main results presented in the letter for the function $\nu_{x}(s)$ describing the late time relaxation of KPZ fluctuations with infinite boundary slopes. Additional checks of the main results against numerical simulations of TASEP are also provided.

\tableofcontents

\vspace{10mm}

\section{Open TASEP}
In this section, we give some details about our numerical simulations for open TASEP, and recall some facts about the generator of the current and the corresponding Perron-Frobenius eigenvector of open TASEP.

\subsection{Simulations}
\label{section simul}
The time complexity of our simulation scales as $\mathcal{O}(L^{5/2}\log L)$ for large $L$. The factor $L^{5/2}$ is the order of magnitude of the total number of time steps, corresponding to one particle moving anywhere in the system, on the KPZ time scale $t\sim L^{3/2}$, noting that $\sim L$ particles move within a unit duration for typical configurations. The remaining $\log L$ factor is needed to read and update the internal hierarchical arrays listing the number of particles that can move in the intervals $i\in[\![2^{k}j,2^{k}(j+1)-1]\!]$ for $k=1,\ldots,K$ and $j=0,\ldots,2^{K-k}-1$ with $L=2^{K}-1$.

For stationary initial condition, with $\alpha=\beta=1$, the initial configuration is generated with rejection sampling, using the combinatorial representation in terms of ``completed configurations'' introduced in \cite{DS2005.1}. An extended configuration with $L$ particles on $2L$ sites is generated uniformly, repeatedly until the condition for completed configurations is satisfied. The algorithm needs $\simeq L/4$ rejections on average, with a time complexity $\mathcal{O}(L)$ for generating each rejected extended configuration. The overall time complexity for sampling the stationary state is then $\mathcal{O}(L^{2})$, which is negligible compared with the time complexity $\mathcal{O}(L^{5/2}\log L)$ needed to run the simulation, both in theory and also in practice for the system sizes $L$ considered.

The simulations were performed with $\alpha=\beta=1$, and averaged respectively over $2^{32},\,2^{30},\,2^{27},\,2^{25},\,2^{22},\,2^{20}$ independent realizations for system size $L=127,\,255,\,511,\,1023,\,2047,\,4095$. For system size $L=1023$, this used for each type of initial condition (stationary, empty, full, flat) around $26000$ hours of computer time on a single core, corresponding to about $70$ hours of real time on the cluster at LPT with typically $384$ cores running in parallel.

\subsection{Generator for the current}
We fix a site $i=0,\ldots,L$ and consider the time-integrated current $Q_{i}(t)$ at position $i$. The joint probability $\mathbb{P}_{t}(C,Q)$ to have at time $t$ the particles organized on the lattice as the configuration $C$, with $Q_{i}(t)=Q$, obeys the master equation
\begin{equation}
	\frac{\dd}{\dd t}\mathbb{P}_{t}(C,Q)=\sum_{C'\neq C}
	\Big(
	w_{C\leftarrow C'}^{+}\,\mathbb{P}_{t}(C',Q-1)
	+w_{C\leftarrow C'}^{0}\,\mathbb{P}_{t}(C',Q)
	-(w_{C'\leftarrow C}^{+}+w_{C'\leftarrow C}^{0})\,\mathbb{P}_{t}(C,Q)
	\Big)\;.
\end{equation}
The summation is over all configurations $C'$ distinct from $C$, with $w_{C\leftarrow C'}^{+}$ (respectively $w_{C\leftarrow C'}^{0}$) equal to the transition rate from $C'$ to $C$ if one can go from $C'$ to $C$ by moving a single particle from site $i$ to $i+1$ (resp. from any site $\ell\neq i$ to $\ell+1$) and $0$ otherwise, with the usual convention that particles moving from site $0$ (respectively $L$) correspond to particles entering (resp. leaving) the system at site $1$ (resp. from site $L$).

Distinct values of $Q$ in the master equation above can be decoupled by considering instead the generating function \cite{DL1998.1} $\mathbb{F}_{t}(C)=\sum_{Q=0}^{\infty}\ee^{\gamma\,Q}\,\mathbb{P}_{t}(C,Q)$. One has
\begin{equation}
	\frac{\dd}{\dd t}\mathbb{F}_{t}(C)=\sum_{C'\neq C}
	\Big(
	(\ee^{\gamma}\,w_{C\leftarrow C'}^{+}+w_{C\leftarrow C'}^{0})\,\mathbb{F}_{t}(C')
	-(w_{C'\leftarrow C}^{+}+w_{C'\leftarrow C}^{0})\,\mathbb{F}_{t}(C)
	\Big)\;.
\end{equation}
Introducing the deformed Markov generator $M_{i}(\gamma)$, with matrix elements $\langle C|M_{i}(\gamma)|C\rangle=-\sum_{C'\neq C}(w_{C'\leftarrow C}^{+}+w_{C'\leftarrow C}^{0})$ and $\langle C|M_{i}(\gamma)|C'\rangle=\ee^{\gamma}\,w_{C\leftarrow C'}^{+}+w_{C\leftarrow C'}^{0}$ for $C\neq C'$, the master equation simply rewrites $\frac{\dd}{\dd t}|\mathbb{F}_{t}\rangle=M_{i}(\gamma)|\mathbb{F}_{t}\rangle$ with $|\mathbb{F}_{t}\rangle=\sum_{C}\mathbb{F}_{t}(C)|C\rangle$. Then, one has
\begin{equation}
	\langle\ee^{\gamma\,Q_{i}(t)}\rangle=\langle\Sigma|\ee^{tM_{i}(\gamma)}|\mathbb{P}_{0}\rangle\;.
\end{equation}
For $\gamma=0$, the generator $M=M_{i}(0)$, independent of $i$, is the usual Markov generator of TASEP, that describes the evolution of configurations only, with no information about the current.

\subsection{Perron-Frobenius stationary eigenstate}
For $\gamma\in\mathbb{R}$ and small enough $\epsilon>0$, all the entries of the matrix $A=\Id+\epsilon\,M_{i}(\gamma)$ are non-negative. Since TASEP is an ergodic Markov process, $M_{i}(\gamma)$ and thus $A$ are irreducible matrices. Calling $\rho$ the spectral radius of $A$ (i.e. the maximum of $|E|$ over all eigenvalues $E$ of $A$), the Perron-Frobenius theorem for non-negative irreducible matrices then implies that $A$ has an eigenvalue equal to $\rho$, which is non-degenerate. Additionally, the corresponding left and right eigenvectors have only positive entries. When $\gamma=0$, these eigenvector are $\langle\Sigma|=(1,\ldots,1)$ and $|\mathbb{P}_{\rm st}\rangle$ the vector of stationary probabilities of TASEP (independent of $i$). In terms of the generator $M_{i}(\gamma)$, the corresponding eigenvalue is the one with largest real part, in a neighbourhood of the line $\gamma\in\mathbb{R}$. The eigenvalue is independent of $i$, and the dependency on $i$ of the eigenvectors follows from $M_{i}(\gamma)=V_{i}^{-1}(\gamma)M_{0}(\gamma)V_{i}(\gamma)$, as
\begin{equation}
	\label{psii[psi]}
	|\psi_{\rm st}^{i}(\gamma)\rangle=V_{i}^{-1}|\psi_{\rm st}^{0}(\gamma)\rangle
	\qquad\text{and}\qquad
	\langle\psi_{\rm st}^{i}(\gamma)|=\langle\psi_{\rm st}^{0}(\gamma)|V_{i}\;.
\end{equation}

\section{Iterative construction of the stationary eigenstate}
In this section, we recall the matrix product representation \cite{GLMV2012.1,L2013.1} for the matrices $T$ and $U$ in the iterative construction (\ref{psi[T,U]}) for the stationary eigenvector. Then, we derive a key height representation for these matrices, which is instrumental in deriving our main result (\ref{nux[h0]}) for the late time correction to stationary large deviations.

\subsection{Matrix product representation}
For general boundary rates $\alpha$, $\beta$, the iterative construction of the stationary vectors $\langle\psi_{\rm st}^{0}(\gamma)|$ and $|\psi_{\rm st}^{0}(\gamma)\rangle$ of $M_{0}(\gamma)$ involves matrices $T(\gamma)$ and $U(\gamma)$ as
\begin{align}
	\label{psi[T,U] alpha beta}
	|\psi_{\rm st}^{0}(\gamma)\rangle&=\big(U(\gamma)T(\gamma)\big)^{n}\,U(0)|\Sigma\rangle+\mathcal{O}(\gamma^{n+1})\nonumber\\
	\langle\psi_{\rm st}^{0}(\gamma)|&=\langle\Sigma|\big(U(\gamma)T(\gamma)\big)^{n}+\mathcal{O}(\gamma^{n+1})\;,
\end{align}
where $U(0)|\Sigma\rangle$ is proportional to the stationary state $|\mathbb{P}_{\rm st}\rangle$. One has the matrix product representation \cite{GLMV2012.1,L2013.1}
\begin{align}
	\label{T[K]}
	\langle C'|T(\gamma)|C\rangle&=\langle W_{T}|A\,K_{n_{1}',n_{1}}\ldots K_{n_{L}',n_{L}}|V_{T}\rangle\\
	\label{U[K]}
	\langle C'|U(\gamma)|C\rangle&=\langle W_{U}|A\,K_{n_{1}',n_{1}}\ldots K_{n_{L}',n_{L}}|V_{U}\rangle\;,
\end{align}
with $n_{i}$, $n_{i}'$ the occupation numbers corresponding to the configurations $C$ and $C'$. The operators $K_{0,0}=1$, $K_{0,1}=E$, $K_{1,0}=D$, $K_{1,1}=1$ act on an infinite dimensional auxiliary space and verify the algebra
\begin{align}
	\label{algebra}
	&DE=1
	&&\langle W_{T}|E=\langle W_{T}|
	&&\langle W_{U}|E=(\alpha^{-1}-1)\,\langle W_{U}|\\
	&AE=\ee^{-\gamma}\,EA
	&&D|V_{T}\rangle=|V_{T}\rangle
	&&D|V_{U}\rangle=(\beta^{-1}-1)\,|V_{U}\rangle\;.\nonumber
\end{align}
With the normalization $\langle W_{T}|V_{T}\rangle=\langle W_{U}|V_{U}\rangle=1$, the algebra above allows to compute in principle all the matrix elements (\ref{T[K]}), (\ref{U[K]}).

We note already that in the limit $\alpha,\beta\to1$, most matrix elements of $U(\gamma)$ vanish, and its remaining non-zero elements are actually independent of $\gamma$, see below.

\subsection{Height representation}
Following \cite{MP2018.1}, we associate with the configurations $C$ and $C'$ in (\ref{T[K]}), (\ref{U[K]}) the height functions $H_{i}$ and $H_{i}'$, $i=0,\ldots,L$ as
\begin{equation}
\label{H[C] SM}
H_{i}=\sum_{\ell=1}^{i}\Big(\frac{1}{2}-n_{\ell}\Big)
\qquad\text{and}\qquad
H_{i}'=\sum_{\ell=1}^{i}\Big(\frac{1}{2}-n_{\ell}'\Big)
\end{equation}
with $n_{i}$, $n_{i}'$ the occupation numbers of the configurations $C$ and $C'$, and consider the path $P_{i}=H_{i}'-H_{i}$, such that $P_{0}=0$ and $P_{i+1}-P_{i}\in\{-1,0,1\}$ for all $i$. Then, we note that $P_{i}=P_{i-1}$ corresponds to $K_{n_{i}',n_{i}}=1$ in (\ref{T[K]}), (\ref{U[K]}), $P_{i}=P_{i-1}+1$ to $K_{n_{i}',n_{i}}=E$ and $P_{i}=P_{i-1}-1$ to $K_{n_{i}',n_{i}}=D$. In order to compute $K_{n_{1}',n_{1}}\ldots K_{n_{L}',n_{L}}$ in (\ref{T[K]}) and (\ref{U[K]}), the constant portions of the path $P_{i}$ can then be erased, as well as the local minima using $DE=1$, see the graphical representation below for an example.
{\setlength{\unitlength}{0.9mm}
	\begin{align}
		&\begin{picture}(100,22)(0,2)
			\put(0,0){\color[rgb]{0.7,0.7,0.7}\line(1,0){60}}
			\put(80,0){\color[rgb]{0.7,0.7,0.7}\line(1,0){20}}
			\put(0,0){\line(1,1){20}}
			\put(20,20){\line(1,0){10}}
			\put(30,20){\line(1,-1){10}}
			\put(40,10){\line(1,0){10}}
			\put(50,10){\line(1,-1){20}}
			\put(70,-10){\line(1,1){20}}
			\put(90,10){\line(1,0){10}}
			\put(10,0){\color[rgb]{0.7,0.7,0.7}\line(0,1){10}}
			\put(20,0){\color[rgb]{0.7,0.7,0.7}\line(0,1){20}}
			\put(30,0){\color[rgb]{0.7,0.7,0.7}\line(0,1){20}}
			\put(40,0){\color[rgb]{0.7,0.7,0.7}\line(0,1){10}}
			\put(50,0){\color[rgb]{0.7,0.7,0.7}\line(0,1){10}}
			\put(90,0){\color[rgb]{0.7,0.7,0.7}\line(0,1){10}}
			\put(100,0){\color[rgb]{0.7,0.7,0.7}\line(0,1){10}}
			\put(3,6){$E$}
			\put(13,16){$E$}
			\put(24,21){$1$}
			\put(34.5,16){$D$}
			\put(44,11){$1$}
			\put(54.5,6){$D$}
			\put(64.5,-4){$D$}
			\put(73,-4){$E$}
			\put(83,6){$E$}
			\put(94,11){$1$}
		\end{picture}
		\quad=\quad
		\begin{picture}(70,22)(0,2)
			\put(0,0){\color[rgb]{0.7,0.7,0.7}\line(1,0){40}}
			\put(60,0){\color[rgb]{0.7,0.7,0.7}\line(1,0){10}}
			\put(0,0){\line(1,1){20}}
			\put(20,20){\line(1,-1){30}}
			\put(50,-10){\line(1,1){20}}
			\put(10,0){\color[rgb]{0.7,0.7,0.7}\line(0,1){10}}
			\put(20,0){\color[rgb]{0.7,0.7,0.7}\line(0,1){20}}
			\put(30,0){\color[rgb]{0.7,0.7,0.7}\line(0,1){10}}
			\put(70,0){\color[rgb]{0.7,0.7,0.7}\line(0,1){10}}
			\put(3,6){$E$}
			\put(13,16){$E$}
			\put(24.5,16){$D$}
			\put(34.5,6){$D$}
			\put(44.5,-4){$D$}
			\put(53,-4){$E$}
			\put(63,6){$E$}
		\end{picture}\nonumber\\
		=&\quad
		\begin{picture}(50,30)(0,2)
			\put(0,0){\color[rgb]{0.7,0.7,0.7}\line(1,0){50}}
			\put(0,0){\line(1,1){20}}
			\put(20,20){\line(1,-1){20}}
			\put(40,0){\line(1,1){10}}
			\put(10,0){\color[rgb]{0.7,0.7,0.7}\line(0,1){10}}
			\put(20,0){\color[rgb]{0.7,0.7,0.7}\line(0,1){20}}
			\put(30,0){\color[rgb]{0.7,0.7,0.7}\line(0,1){10}}
			\put(50,0){\color[rgb]{0.7,0.7,0.7}\line(0,1){10}}
			\put(3,6){$E$}
			\put(13,16){$E$}
			\put(24.5,16){$D$}
			\put(34.5,6){$D$}
			\put(43,6){$E$}
		\end{picture}
		\quad=\quad
		\begin{picture}(30,25)(0,2)
			\put(0,0){\color[rgb]{0.7,0.7,0.7}\line(1,0){30}}
			\put(0,0){\line(1,1){20}}
			\put(20,20){\line(1,-1){10}}
			\put(10,0){\color[rgb]{0.7,0.7,0.7}\line(0,1){10}}
			\put(20,0){\color[rgb]{0.7,0.7,0.7}\line(0,1){20}}
			\put(30,0){\color[rgb]{0.7,0.7,0.7}\line(0,1){10}}
			\put(3,6){$E$}
			\put(13,16){$E$}
			\put(24.5,16){$D$}
		\end{picture}\quad,\nonumber\\[-4mm]\nonumber
\end{align}}\noindent
where each picture represents the product of the operators above each section of the path, ordered from left to right. Noting that the height of the path on the left, on the right, as well as its maximal value are preserved by the procedure outlined above, we finally obtain the identity
\begin{equation}
	K_{n_{1}',n_{1}}\ldots K_{n_{L}',n_{L}}=E^{\max(H'-H)}D^{\max(H'-H)+N'-N}\;,
\end{equation}
with $N=\sum_{\ell=1}^{L}n_{\ell}$, $N'=\sum_{\ell=1}^{L}n_{\ell}'$ the total number of particles in the configurations $C$, $C'$, and the maxima computed over all the sites $i=0,\ldots,L$. The exponent of $E$ (respectively $D$) is non-negative since $H_{0}=H_{0}'=0$ (resp. $H_{L}=\frac{L}{2}-N$ and $H_{L}'=\frac{L}{2}-N'$). The matrix product representations (\ref{T[K]}) and (\ref{U[K]}) can finally be computed using the rest of the algebra (\ref{algebra}), leading for arbitrary $\alpha$, $\beta$ to
\begin{equation}
\label{T[H] SM}
\langle C^{2}|T(\gamma)|C^{1}\rangle=\ee^{-\gamma\,\max(H^{2}-H^{1})}\;,
\end{equation}
and
\begin{equation}
\label{U[H] alpha beta}
\langle C^{2}|U(\gamma)|C^{1}\rangle=\frac{(\alpha^{-1}-1)^{\max(H^{2}-H^{1})}\,(\beta^{-1}-1)^{\max(H^{2}-H^{1})+N^{2}-N^{1}}}{\ee^{\gamma\,\max(H^{2}-H^{1})}}\;,
\end{equation}
with $N^{1}$ and $N^{2}$ the total number of particles in the system for the configurations $C^{1}$ and $C^{2}$. We specialize from now on to the special case $\alpha=\beta=1$. Taking the limit $\alpha,\beta\to1$ in (\ref{U[H] alpha beta}), $U(\gamma)$ reduces in particular to
\begin{equation}
\label{U[H] inf inf}
\langle C^{2}|U(\gamma)|C^{1}\rangle=1_{\{H_{L}^{1}=H_{L}^{2}\}}\,1_{\{\forall i=0,\ldots,L:H_{i}^{1}\geq H_{i}^{2}\}}\;.
\end{equation}
All the non-zero matrix elements $\langle C^{2}|U(\gamma)|C^{1}\rangle$ have $\max(H^{2}-H^{1})=0$, which explains why $U(\gamma)$ is then independent of $\gamma$.

\section{KPZ fluctuations}
In this section, we give a detailed derivation of our main result (\ref{nux[h0]}) for the late time approach to stationary large deviations at the KPZ fixed point with infinite boundary slopes. We also write down explicit expressions for the corresponding first cumulants, which are checked against additional numerical simulations.

\subsection{Stationary cumulants}
The first stationary cumulants $c_{k}$, defined in (\ref{cumulants[ck,dk]}), can be extracted from $\mu(s)=s+\chi(v)/2$, $s=\chi'(v)$ by expanding
\begin{equation}
\label{chi SM}
\chi(v)=-\frac{1}{4\sqrt{\pi}}\sum_{k=1}^{\infty}\frac{(2k)!}{k^{k+5/2}\,k!}\,\Big(-\frac{\ee^{v}}{4}\Big)^{k}\;.
\end{equation}
in powers of $\ee^{v}$ \cite{GLMV2012.1}. One has
\begin{align}
	c_{1}&=3/2\\
	c_{2}&=\frac{3\sqrt{\pi}}{4\sqrt{2}}\approx0.939986\\
	c_{3}&=\Big(\frac{27}{8}-\frac{160}{27\sqrt{3}}\Big)\,\pi\approx-0.145565\\
	c_{4}&=\Big(\frac{945}{32}-\frac{80\sqrt{2}}{\sqrt{3}}+\frac{405}{8\sqrt{2}}\Big)\,\pi^{3/2}\approx0.049025\;.
\end{align}

\subsection{Correction $\nu_{x}(s)$ to stationary large deviations}
From (\ref{psii[psi]}) and (\ref{psi[T,U] alpha beta}),
\begin{equation}
	\frac{\langle\Sigma|\psi_{\rm st}^{i}(\gamma)\rangle\langle\psi_{\rm st}^{i}(\gamma)|\mathbb{P}_{0}\rangle}{\langle\psi_{\rm st}^{i}(\gamma)|\psi_{\rm st}^{i}(\gamma)\rangle}
	=\frac{\langle\Sigma|V_{i}^{-1}(U(\gamma)T(\gamma))^{n}U(0)|\Sigma\rangle\;\langle\Sigma|(U(\gamma)T(\gamma))^{n}V_{i}|\mathbb{P}_{0}\rangle}{\langle\Sigma|(U(\gamma)T(\gamma))^{2n}U(0)|\Sigma\rangle}
	+\mathcal{O}(\gamma^{n+1})\;,
\end{equation}
where the dependency in $i$ cancels in the denominator. Inserting the resolution of the identity $1=\sum_{C}|C\rangle\langle C|$, with a sum over all $2^{L}$ configurations of the particles, and using $\langle\Sigma|C\rangle=\langle C|\Sigma\rangle=1$ for any configuration $C$, one has
\begin{align}
	&\frac{\langle\Sigma|\psi_{\rm st}^{i}(\gamma)\rangle\langle\psi_{\rm st}^{i}(\gamma)|\mathbb{P}_{0}\rangle}{\langle\psi_{\rm st}^{i}(\gamma)|\psi_{\rm st}^{i}(\gamma)\rangle}
	=\bigg(\sum_{C^{0},\ldots,C^{2n}}\langle C^{2n}|U(\gamma)|C^{2n-1}\rangle\langle C^{2}|U(\gamma)|C^{1}\rangle\langle C^{1}|T(\gamma)|C^{0}\rangle\langle C^{0}|V_{i}|\mathbb{P}_{0}\rangle\bigg)\\
	&\qquad\times\frac{\sum_{C^{1},\ldots,C^{2n+2}}\langle C^{2n+2}|V_{i}^{-1}(\gamma)|C^{2n+2}\rangle\langle C^{2n+2}|U(\gamma)|C^{2n+1}\rangle\ldots\langle C^{4}|U(\gamma)|C^{3}\rangle\langle C^{3}|T(\gamma)|C^{2}\rangle\langle C^{2}|U(0)|C^{1}\rangle}{\sum_{C^{1},\ldots,C^{4n+2}}\langle C^{4n+2}|U(\gamma)|C^{4n+1}\rangle\ldots\langle C^{4}|U(\gamma)|C^{3}\rangle\langle C^{3}|T(\gamma)|C^{2}\rangle\langle C^{2}|U(0)|C^{1}\rangle}\nonumber\\
	&\qquad+\mathcal{O}(\gamma^{n+1})\;.\nonumber
\end{align}
The matrix elements of $T(\gamma)$ are given in (\ref{T[H] SM}) in terms of the corresponding discrete height functions from (\ref{H[C] SM}). At large $L$, the KPZ fixed point with infinite boundary slopes is expected in the whole maximal current phase $\alpha,\beta>1/2$ (fine-tuning $\alpha-1/2\sim\beta-1/2\sim L^{-1/2}$ would be required for finite boundary slopes). For simplicity, we set $\alpha=\beta=1$, so that the matrix elements of $U(\gamma)$ are given by (\ref{U[H] inf inf}). Then, Donsker's theorem (\ref{Donsker}) leads at large $L$ with $i=xL$, $\gamma$ scaling as in (\ref{E[mu] ; gamma[s]}) and an initial condition of the form (\ref{H0[h0]}) to the following Brownian average
\begin{equation}
	\frac{\langle\Sigma|\psi_{\rm st}^{i}(\gamma)\rangle\langle\psi_{\rm st}^{i}(\gamma)|\mathbb{P}_{0}\rangle}{\langle\psi_{\rm st}^{i}(\gamma)|\psi_{\rm st}^{i}(\gamma)\rangle}\simeq
	\frac{\langle\ee^{sw_{2n+2}(x)-s\sum_{k=1}^{n}\max(w_{2k+1}-w_{2k})}\rangle\;\langle\ee^{-sh_{0}(x)-s\max(w_{1}-h_{0})-s\sum_{k=1}^{n-1}\max(w_{2k+1}-w_{2k})}\rangle}{\langle\ee^{-s\sum_{k=1}^{2n}\max(w_{2k+1}-w_{2k})}\rangle}
	+\mathcal{O}(s^{n+1})\;,
\end{equation}
where everywhere, the independent Wiener processes $w_{j}$ are conditioned for all $k$ on $w_{2k}(1)=w_{2k-1}(1)$ and $w_{2k}(u)\leq w_{2k-1}(u)$, $u\in[0,1]$.

Setting $p_{k}^{-}=w_{2k}$ and $p_{k}^{+}=w_{2k-1}$, the conditioning on $p_{k}^{-}(1)=p_{k}^{+}(1)$ and $p_{k}^{-}(u)\leq p_{k}^{+}(u)$, $u\in[0,1]$ can be written alternatively in terms of Wiener processes $\tilde{w}_{k}$ and Brownian excursions $\tilde{e}_{k}$ \emph{independent of each other} as
\begin{equation}
	p_{k}^{\pm}=\frac{\tilde{w}_{k}-\tilde{e}_{k}}{\sqrt{2}}\;,
\end{equation}
see section~\ref{section equivalence w e}. Then, for deterministic initial condition associated with the discrete height function $H^{0}$, (\ref{Q[H]}) leads to
\begin{equation}
	\langle\ee^{\gamma\,H_{i}(t)}\rangle
	=\langle\ee^{\gamma\,(Q_{i}(t)+H_{i}^{0})}\rangle
	=\ee^{\gamma H_{i}^{0}}\langle\Sigma|\ee^{tM_{i}(\gamma)}|\mathbb{P}_{0}\rangle
	\underset{t\to\infty}{\simeq}\ee^{\gamma H_{i}^{0}}\frac{\langle\Sigma|\psi_{\rm st}^{i}(\gamma)\rangle\langle\psi_{\rm st}^{i}(\gamma)|\mathbb{P}_{0}\rangle}{\langle\psi_{\rm st}^{i}(\gamma)|\psi_{\rm st}^{i}(\gamma)\rangle}\,\ee^{tE_{\rm st}(\gamma)}\;.
\end{equation}
Using (\ref{GF[h]}), (\ref{H[h]}), (\ref{E[mu] ; gamma[s]}) and the scaling in (\ref{t[tau]}) for $t$, we obtain the expression (\ref{nux[h0]}) for $\ee^{\nu_{x}(s)}$, which remains valid for random initial condition by averaging over $h_{0}$.

We note that the dependency on $x$ of $\nu_{x}(s)$ is contained solely in the factor $Z_{n}(s;x)$: at late times, the spatial height increments $h(x,\tau)-h(0,\tau)$ are thus independent of the initial condition, as expected physically. Note however that the random variables $h(x,\tau)-h(0,\tau)$ and $h(0,\tau)$ remain coupled even at late times, so that the dependency in $x$ of $\nu_{x}(s)$ can not be computed simply from the fact that $h(x,\tau)-h(0,\tau)$ has the same statistics as the stationary state $h_{\rm st}(x)$ when $\tau\to\infty$. In the driven particle language, this means that the time-integrated current and the density are still correlated at late times, leading to the well known phenomenon that conditioning on atypical values of the current induces changes in the effective interaction between particles.

Additionally, from the conjectured exact expression (\ref{omega stat}) for $\omega_{\rm stat}$ obtained from Bethe ansatz numerics, the function $\nu_{0}(s)$ for stationary initial condition has the alternative explicit expression (\ref{nu0 stat}), see section~\ref{section Nst asymptotics} for a detailed derivation.

\subsection{Initial conditions corresponding to macroscopic density profiles}
One can consider more generally initial conditions corresponding to a macroscopic density profile $\rho_{0}(x)$, such that the number of particles initially within the $L\,\dd x$ sites around $i=x L$ is equal to $L\,\rho_{0}(x)\,\dd x$ for $1\ll L\,\dd x\ll L$. This type of initial condition does not correspond to a small perturbation of order $L^{-1/2}$ over the flat profile $\rho=1/2$ as in (\ref{H0[h0]}). On the Euler scale $t\sim L$, deterministic hydrodynamics leads to a relaxation to the constant density profile $\rho=1/2$. Then, on the KPZ time scale (\ref{t[tau]}), the KPZ height $h(x,\tau)$ converges for small $\tau$ to a sum of narrow wedges $\frac{|x-x_{0}|}{0}-\infty$ centered at the global minima $x_{0}$ of the initial height $\int_{0}^{x}(\frac{1}{2}-\rho_{0}(u))\dd u$, acting as a singular initial condition for the KPZ height. The connection (\ref{H[h]}) between the discrete TASEP height and the KPZ height is then modified as
\begin{equation}
	\label{H[h,D]}
	H_{i}(t)\simeq\frac{t}{4}+D\,L+\frac{\sqrt{L}}{2}\,h(x,\tau)\;,
\end{equation}
where the constant $D=\min_{x}\int_{0}^{x}(\frac{1}{2}-\rho_{0}(u))\dd u$ is the deterministic contribution of the hydrodynamic evolution on the Euler scale, see e.g. \cite{P2015.3}.

We focus in the following on two simple cases, corresponding to a system initially completely empty of filled with particles, for which one has respectively $D_{\rm empty}=0$ and $D_{\rm full}=-1/2$, and corresponding to a (half) narrow wedge located respectively at $x_{0}=0$ and $x_{0}=1$. A similar computation as before leads in both cases to
\begin{equation}
	\label{nux[h0 nw]}
	\ee^{\nu_{x}^{\rm nw}(s)}=\,\frac{Z_{n}(s;x)}{Z_{2n}(s;0)}\,\Big\langle\ee^{s\sum\limits_{k=1}^{n-1}\min(p_{k}^{-}-p_{k+1}^{+})}\Big\rangle
	+\mathcal{O}(s^{n+1})
\end{equation}
with $Z_{m}(s;x)$ as below (\ref{nux[h0]}).

Additionally, from the conjectured exact expression (\ref{omega empty}) and below for empty and full initial condition, the function $\nu_{0}(s)$ is found to have the alternative explicit expression (\ref{nu0 nw}) for both narrow wedge initial conditions above, see section~\ref{section Nst asymptotics} for a detailed derivation.

\subsection{Late time correction to the stationary cumulants of $h(x,\tau)$}
The coefficients $d_{1}(x)$ and $d_{2}(x)$ in (\ref{cumulants[ck,dk]}) can be extracted from the expansion in powers of $s$ of the expression (\ref{nux[h0]}) for $\ee^{\nu_{0}(s)}$ with $n=2$. Several Brownian averages then need to be computed, see section~\ref{section Brownian averages}. Using (\ref{<w>}), (\ref{<e>}) and the Brownian averages involving maxima in table~\ref{table Brownian averages} (and in some places the fact that the opposite of a Wiener process is also a Wiener process), one finds (\ref{d1[h0]}) for $d_{1}(x)$ and
\begin{align}
	\label{d2[h0]}
	d_{2}(x)=4&+\frac{45\pi}{8}-\frac{320\pi}{27\sqrt{3}}+2x-\frac{3x^{2}}{2}-\frac{6\sqrt{x(1-x)}}{\sqrt{2}}-\frac{4x(1-x)}{\pi}+\big\langle p_{2}^{-}(x)\,\min(p_{1}^{-}-p_{2}^{+})\big\rangle\nonumber\\
	&+\mathrm{Var}\big(\min(h_{0}+h_{\rm st})\big)
	+\frac{3\sqrt{\pi}}{\sqrt{2}}\big\langle\min(h_{0}+h_{\rm st})\big\rangle
	+2\big\langle\min(h_{0}-p_{1}^{+})\min(p_{1}^{-}-p_{2}^{+})\big\rangle\;.
\end{align}

\begin{figure}
	\begin{tabular}{lll}
		\hspace*{-3mm}
		\begin{tabular}{l}
			\includegraphics[width=86mm]{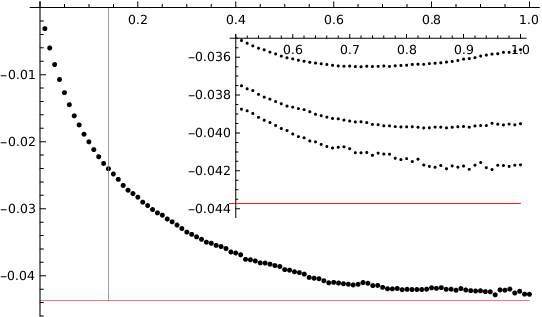}
		\end{tabular}
		\begin{picture}(0,0)
			\put(-3,26.5){$\tau$}
			\put(-71,26.5){$\tau_{*}$}
			\put(-30,-13){$\langle h(0,\tau)^{3}\rangle_{\rm c}-c_{3}\,\tau$}
		\end{picture}
		&&
		\begin{tabular}{l}
			\includegraphics[width=86mm]{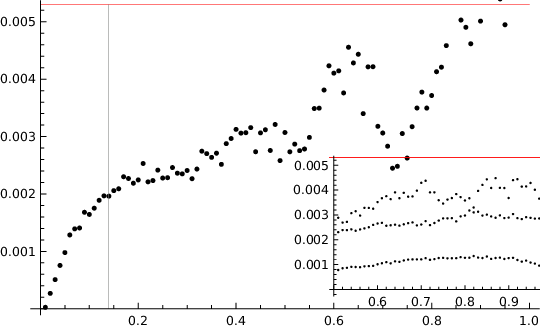}
		\end{tabular}
		\begin{picture}(0,0)
			\put(-3,-20.7){$\tau$}
			\put(-72,-24){$\tau_{*}$}
			\put(-65,20){$\langle h(0,\tau)^{4}\rangle_{\rm c}-c_{4}\,\tau$}
		\end{picture}
	\end{tabular}
	\caption{Third and fourth cumulants of $h(0,\tau)$ plotted as a function of time for infinite boundary slopes and stationary initial condition. The data were obtained from simulations of open TASEP with $L=1023$ sites, boundary rates $\alpha=\beta=1$. The red, horizontal lines are the exact expression in (\ref{d3(0) stat}) and (\ref{d4(0) stat}) for $d_{3}(0)$ and $d_{4}(0)$. As seen in the inset, which represents simulations with $L=127,255,511$ (from top to bottom on the left, from bottom to top on the right), finite size corrections $\sim1/L$ are present. The simulations were averaged over a number of independent realizations detailed in section~\ref{section simul}.}
	\label{fig simul TASEP stat}
\end{figure}
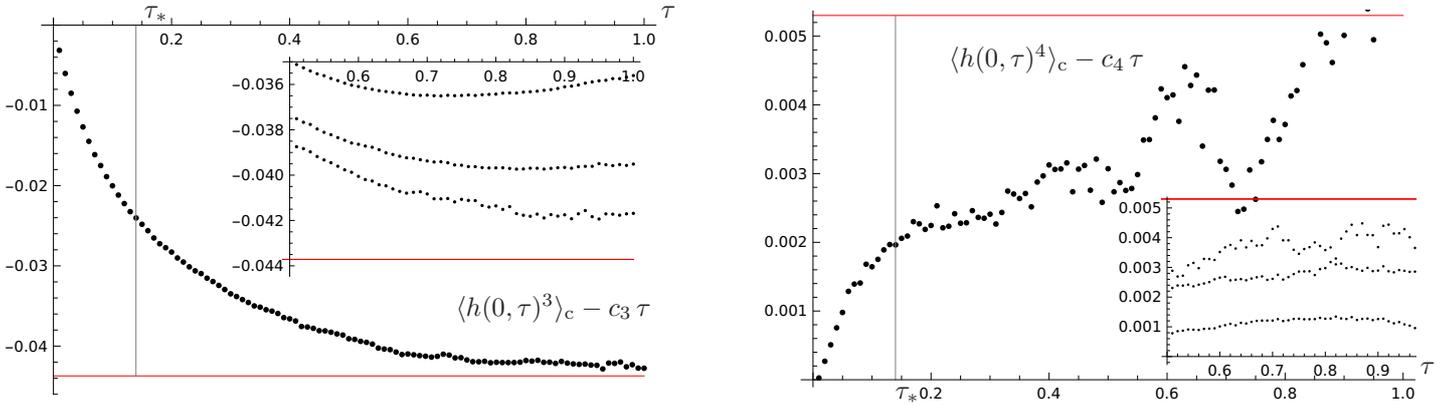

Explicit expressions can be obtained for specific initial conditions. For stationary initial condition $h_{0}=h_{\rm st}$, using again the results in table~\ref{table Brownian averages}, (\ref{d1[h0]}) and (\ref{d2[h0]}) give
\begin{align}
	d_{1}^{\rm stat}(0)&=0\\
	d_{2}^{\rm stat}(0)&=4+\frac{9\pi}{4}-\frac{160\pi}{27\sqrt{3}}\approx0.320143\;,
\end{align}
while the exact expression (\ref{nu0 stat}) for $\nu_{0}(s)$ additionally leads to
\begin{align}
	\label{d3(0) stat}
	d_{3}^{\rm stat}(0)&=\Big(\frac{315}{8}-\frac{80\sqrt{2}}{\sqrt{3}}+\frac{81}{2\sqrt{2}}\Big)\,\pi^{3/2}-6\sqrt{2\pi}\approx-0.0437157\\
	\label{d4(0) stat}
	d_{4}^{\rm stat}(0)&=\Big(\frac{1280}{9 \sqrt{3}}-54\Big)\,\pi+\Big(\frac{9642863}{11664}+\frac{2835}{2 \sqrt{2}}-\frac{2320}{\sqrt{3}}-\frac{3483648}{3125\sqrt{5}}\Big)\,\pi^2\approx0.00530178\;.
\end{align}
These results are in good agreement with simulations of TASEP, see figures~\ref{fig simul var TASEP} and \ref{fig simul TASEP stat}.

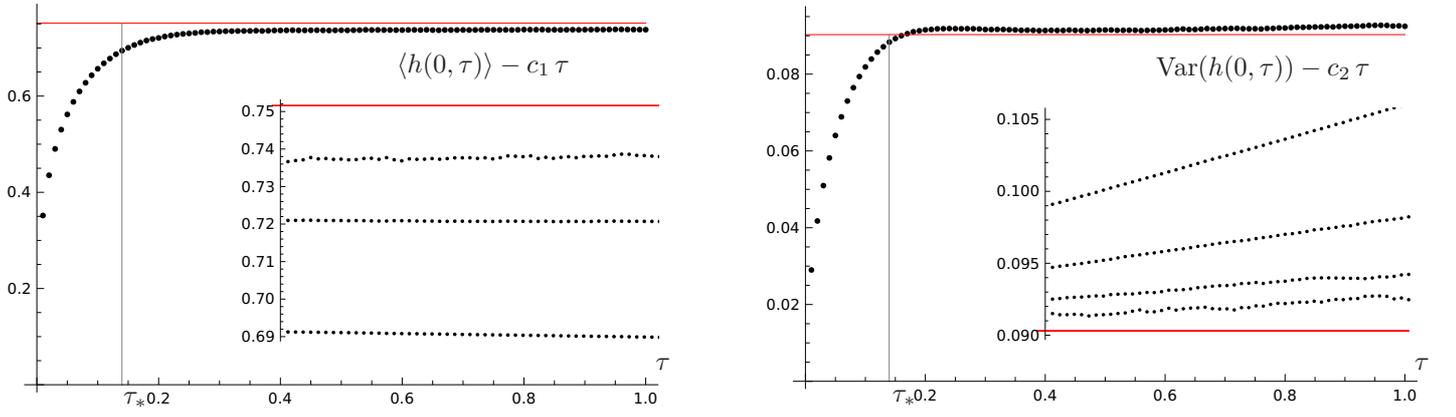
\begin{figure}
	\begin{tabular}{lll}
		\hspace*{-3mm}
		\begin{tabular}{l}
			\includegraphics[width=86mm]{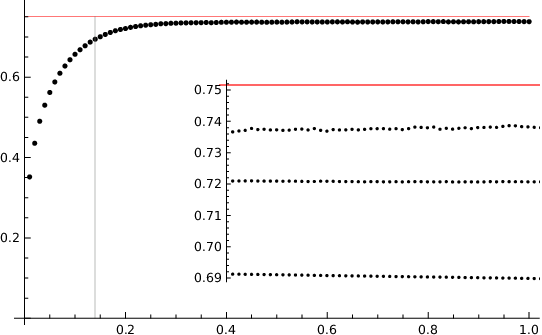}
		\end{tabular}
		\begin{picture}(0,0)
			\put(-4,-20.5){$\tau$}
			\put(-74,-25){$\tau_{*}$}
			\put(-38,19){$\langle h(0,\tau)\rangle-c_{1}\,\tau$}
		\end{picture}
		&&
		\begin{tabular}{l}
			\includegraphics[width=86mm]{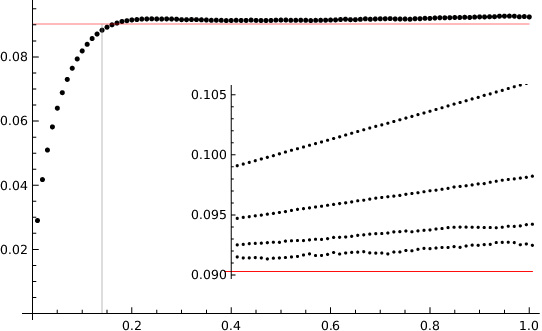}
		\end{tabular}
		\begin{picture}(0,0)
			\put(-4,-20.5){$\tau$}
			\put(-72.7,-24.7){$\tau_{*}$}
			\put(-38,18.5){$\mathrm{Var}(h(0,\tau))-c_{2}\,\tau$}
		\end{picture}
	\end{tabular}
	\vspace{-3.5mm}\\
	\caption{Two first cumulants of $h(0,\tau)$ plotted as a function of time for infinite boundary slopes and flat initial condition $h_{0}(x)=0$. The average (respectively variance) of $h(0,\tau)$ is displayed on the left (resp. right). The data were obtained from simulations of open TASEP with $L=4095$ (resp. $L=1023$) sites, prepared initially with $(L-1)/2$ particles located at the even sites, with boundary rates $\alpha=\beta=1$. The red, horizontal line is the exact expression (\ref{d1(0) flat}) (resp. the numerical value (\ref{d2(0) flat})). As seen in the inset, which represents from bottom to top simulations with $L=255,1023,4095$ (resp. from top to bottom with $L=127,255,511,1023$), the small discrepancy observed at late times is consistent with a finite size correction $\sim1/\sqrt{L}$ (resp. $\sim1/L$). The simulations were averaged over a number of independent realizations detailed in section~\ref{section simul}.}
	\label{fig simul TASEP flat}
\end{figure}

For flat initial condition $h_{0}=0$, one finds from (\ref{d1[h0]}) and (\ref{d2[h0]})
\begin{equation}
	\label{d1(0) flat}
	d_{1}^{\rm flat}(0)=\frac{3\sqrt{\pi}}{2\sqrt{2}}-\frac{2}{\sqrt{\pi}}\approx0.751592
\end{equation}
and
\begin{equation}
	\label{d2(0) flat}
	d_{2}^{\rm flat}(0)=\frac{11}{2}-3\sqrt{2}-\frac{4}{\pi}+\frac{45\pi}{8}-\frac{320\pi}{27\sqrt{3}}+\langle\max(e_{1}+w_{1})\max(e_{1}+e_{2}-w_{1}-w_{2})\rangle\approx0.0903\;.
\end{equation}
These results are again in good agreement with simulations of TASEP, see figure~\ref{fig simul TASEP flat}.

\begin{figure}
	\begin{tabular}{lll}
		\hspace*{-3mm}
		\begin{tabular}{l}
			\includegraphics[width=86mm]{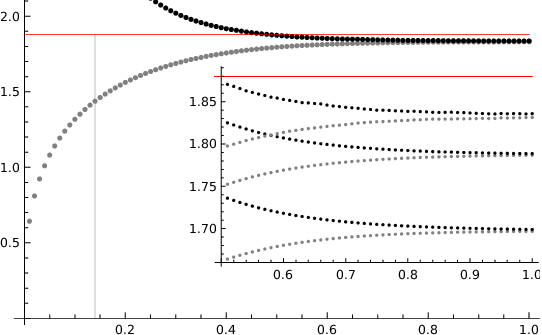}
		\end{tabular}
		\begin{picture}(0,0)
			\put(-4,-21){$\tau$}
			\put(-74,-25.2){$\tau_{*}$}
			\put(-30,26){$\langle h(0,\tau)\rangle-c_{1}\,\tau$}
		\end{picture}
		&&
		\begin{tabular}{l}
			\includegraphics[width=86mm]{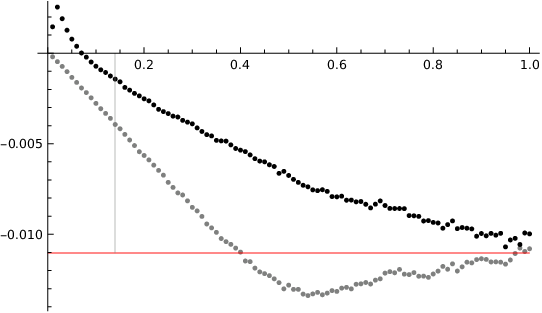}
		\end{tabular}
		\begin{picture}(0,0)
			\put(-4,19.5){$\tau$}
			\put(-70,19){$\tau_{*}$}
			\put(-35,7){$\langle h(0,\tau)^{3}\rangle_{\rm c}-c_{3}\,\tau$}
		\end{picture}
	\end{tabular}
	\caption{Average and third cumulant of the KPZ height $h(0,\tau)$, defined from TASEP as in (\ref{H[h,D]}), with boundary rates $\alpha=\beta=1$, for empty (grey dots) and full (black dots) initial condition, plotted as a function of time $\tau$ as in (\ref{t[tau]}). The red, horizontal line is the exact expression in (\ref{d1(0) nw}), (\ref{d3(0) nw}). Left: $\langle h(0,\tau)\rangle-c_{1}\,\tau$ with $L=4095$ (inset: $L=255,1023,4095$ from bottom to top, indicating a finite size correction $\sim L^{-1/2}$). Right: $\langle h(0,\tau)^{3}\rangle_{\rm c}-c_{3}\,\tau$ with $L=1023$. The simulations were averaged over a number of independent realizations detailed in section~\ref{section simul}.}
	\label{fig simul TASEP nw}
\end{figure}
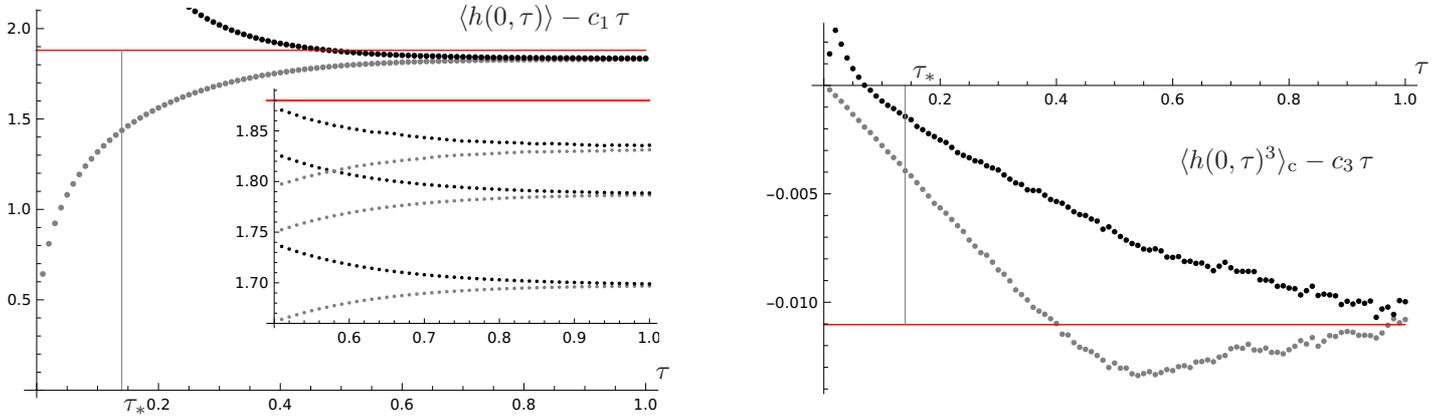

For narrow wedge initial condition located at position $x_{0}=0$ or $x_{0}=1$, (\ref{nux[h0 nw]}) leads to
\begin{align}
	\label{d1(0) nw}
	d_{1}^{\rm nw}(0)&=\frac{3\sqrt{\pi}}{2\sqrt{2}}\approx1.87997\\
	\label{d2(0) nw}
	d_{2}^{\rm nw}(0)&=4+\frac{45\pi}{8}-\frac{320\pi}{27\sqrt{3}}\approx0.174577\;.
\end{align}
This is in agreement with the conjectured exact formula (\ref{nu0 nw}) for $\nu_{0}(s)$, which additionally leads to
\begin{align}
	\label{d3(0) nw}
	d_{3}^{\rm nw}(0)&=\Big(\frac{945}{16}-\frac{400\sqrt{2}}{3\sqrt{3}}+\frac{297}{4\sqrt{2}}\Big)\pi^{3/2}-6\sqrt{2\pi}\approx-0.011032\\
	\label{d4(0) nw}
	d_{4}^{\rm nw}(0)&=\Big(\frac{1280}{9\sqrt{3}}-54\Big)\pi +\Big(\frac{29581871}{23328}+\frac{16065}{8 \sqrt{2}}-\frac{3520}{\sqrt{3}}-\frac{4644864}{3125\sqrt{5}}\Big)\pi^{2}\approx-0.0030518\;.
\end{align}
These results are in good agreement with simulations of TASEP with empty and full initial conditions, see figure~\ref{fig simul TASEP nw}.

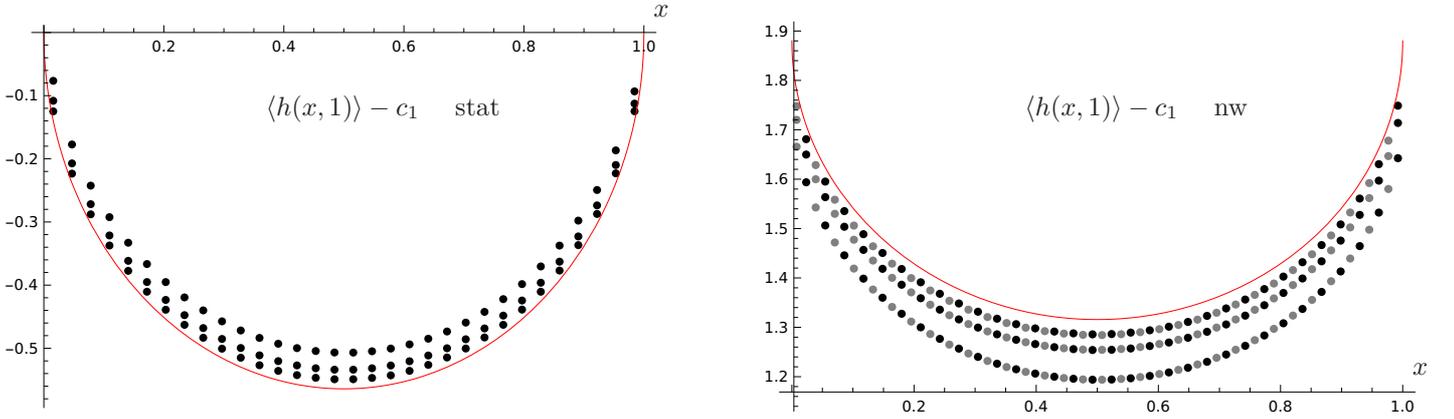
\begin{figure}
	\begin{tabular}{lll}
		\hspace*{-3mm}
		\begin{tabular}{l}
			\includegraphics[width=86mm]{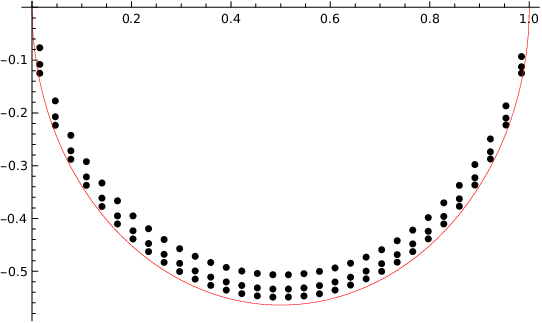}
		\end{tabular}
		\begin{picture}(0,0)
			\put(-4,28){$x$}
			\put(-55,15){$\langle h(x,1)\rangle-c_{1}$ \quad stat}
		\end{picture}
		&&
		\begin{tabular}{l}
			\includegraphics[width=86mm]{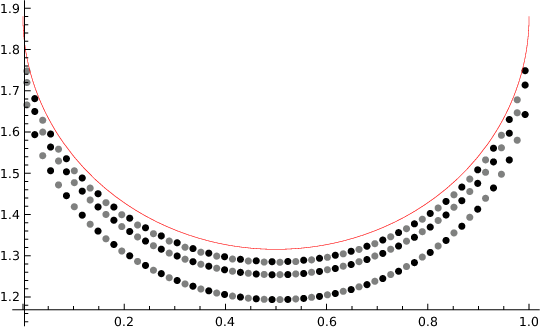}
		\end{tabular}
		\begin{picture}(0,0)
			\put(-4,-19.5){$x$}
			\put(-55,15){$\langle h(x,1)\rangle-c_{1}$ \quad nw}
		\end{picture}
	\end{tabular}
	\caption{Average of the KPZ height $\langle h(x,\tau)\rangle-c_{1}\tau$ at time $\tau=1\gg\tau_{*}$ plotted as a function of the position $x$, defined from TASEP with boundary rates $\alpha=\beta=1$ for stationary (left), empty (right, grey dots) and full (right, black dots) initial conditions. The dots correspond to simulations with $L=255,1023,4095$ (left: from top to bottom; right: from bottom to top), and indicate a finite size correction $\sim L^{-1/2}$. In each case, the red curve is the exact expression from (\ref{d1(x)-d1(0)}). The simulations were averaged over a number of independent realizations detailed in section~\ref{section simul}.}
	\label{fig simul TASEP nw x}
\end{figure}

\begin{figure}
	\begin{tabular}{l}
		\includegraphics[width=86mm]{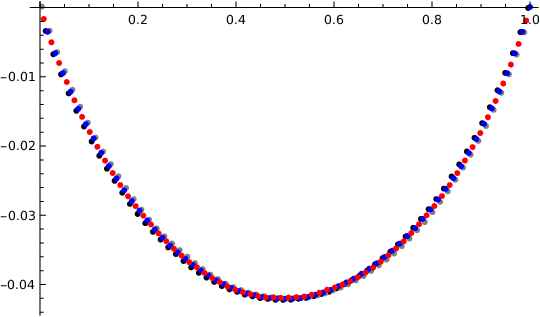}
		\begin{picture}(0,0)
			\put(-1,50){$x$}
			\put(-71,35){$\displaystyle\mathrm{Var}(h(x,1))-\frac{\mathrm{Var}(h(0,1))+\mathrm{Var}(h(1,1))}{2}$}
		\end{picture}
	\end{tabular}
	\caption{Shifted variance of the KPZ height $\mathrm{Var}(h(x,1))-\frac{\mathrm{Var}(h(0,1))+\mathrm{Var}(h(1,1))}{2}$ at time $\tau=1\gg\tau_{*}$ plotted as a function of the position $x$, defined from TASEP with $L=1023$ and boundary rates $\alpha=\beta=1$ for stationary (blue dots), flat (green; even sites occupied), empty (grey dots) and full (black dots) initial conditions. The simulations were averaged over a number of independent realizations detailed in section~\ref{section simul}.}
	\label{fig simul TASEP Var x}
\end{figure}
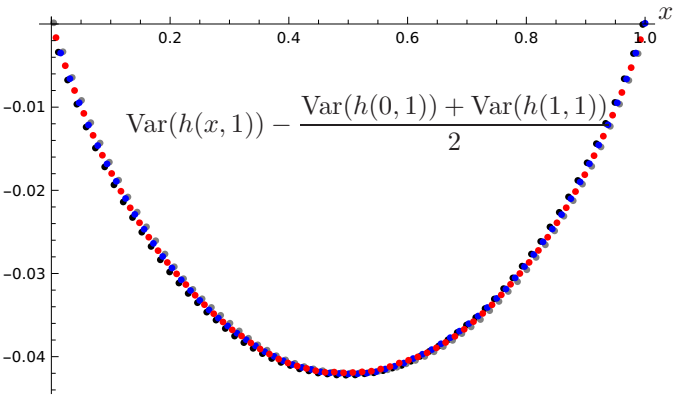

Finally, we emphasize that the dependency in $x$ of the cumulants $d_{k}(x)$ is independent of the initial condition, see figure~\ref{fig simul TASEP Var x} for the variance from simulations of TASEP. In particular, (\ref{d1[h0]}) implies
\begin{equation}
	\label{d1(x)-d1(0)}
	d_{1}(x)-d_{1}(0)=-\frac{2\sqrt{x(1-x)}}{\sqrt{\pi}}
\end{equation}
always. This is in agreement with simulations of TASEP, see figure~\ref{fig simul TASEP nw x}.

\section{Computations with Wiener processes}
In this section, we detail standard calculations with Wiener processes that are used in various places in the letter.

\subsection{Brownian excursions}
Brownian excursions $e(x)$, $x\in[0,1]$, are Wiener processes conditioned on $e(0)=e(1)=0$ and $e(x)\geq0$ for all $x$. There is a well known alternative representation of Brownian excursions, with the same joint probability distribution, in terms of non-conditioned Wiener processes. This representation, described below, is particularly useful for numerical simulations of Brownian excursions.

First, one constructs a Brownian bridge $b(x)$, $x\in[0,1]$, i.e. a Wiener process conditioned on $b(0)=b(1)=0$, using the simple alternative representation as
\begin{equation}
	b(x)=w(x)-xw(1)
\end{equation}
with $w(x)$ a Wiener process. Then, one has
\begin{align}
	e(x)=b(x_{m}+x)-b(x_{m})&\qquad\text{for}\;x\in[0,1-x_{m}]\nonumber\\
	e(x)=b(x_{m}+x-1)-b(x_{m})&\qquad\text{for}\;x\in[1-x_{m},1]\;,
\end{align}
where $x_{m}$ is any point where $b(x)$ reaches its minimum.

\subsection{Non-intersecting Wiener processes and Brownian excursions}
\label{section equivalence w e}
We consider two independent Wiener processes $w_{1}$ and $w_{2}$, conditioned on the event $w_{1}(1)=w_{2}(1)$ and $w_{1}(u)\geq w_{2}(u)$ for all $u\in[0,1]$, understood as the limit $\epsilon,\delta\to0^{+}$ of the conditioning on the finite probability event $w_{1}(1)-w_{2}(1)\in[0,\epsilon]$ and $w_{1}(u)-w_{2}(u)\geq-\delta$ for all $u\in[0,1]$. We show in this section that $\frac{w_{1}+w_{2}}{\sqrt{2}}$ and $\frac{w_{1}-w_{2}}{\sqrt{2}}$ are respectively a Wiener process and a Brownian excursion, independent of each other.

Using repeatedly the Markov property of the Wiener process, the joint probability density of $w_{1}$ and $w_{2}$ at $n$ points $0<x_{1}<\ldots<x_{n}<1$ (with $\vec{x}=(x_{1},\ldots,x_{n})$, etc.)
\begin{equation}
	p(\vec{x},\vec{y},\vec{z})=\frac{\mathbb{P}(\forall j=1,\ldots,n:\;w_{1}(x_{j})\in[y_{j},y_{j}+\dd y_{j}]\;\text{and}\;w_{2}(x_{j})\in[z_{j},z_{j}+\dd z_{j}])}{\prod_{j=1}^{n}\dd y_{j}\dd z_{j}}
\end{equation}
is expressed after standard calculations as
\begin{align}
	p(\vec{x},\vec{y},\vec{z})=\Big(\prod_{j=1}^{n}1_{\{y_{j}>z_{j}\}}\Big)\,
	\frac{(y_{1}-z_{1})\ee^{-\frac{y_{1}^{2}}{2x_{1}}-\frac{z_{1}^{2}}{2x_{1}}}}{2\pi x_{1}^{2}}\,&
	\frac{(y_{n}-z_{n})\ee^{\frac{(y_{n}+z_{n})^{2}}{4(1-x_{n})}-\frac{y_{n}^{2}}{2(1-x_{n})}-\frac{z_{n}^{2}}{2(1-x_{n})}}}{(1-x_{n})^{3/2}}\\
	&\times\prod_{j=2}^{n}\det\bigg(\begin{array}{cc}G_{x_{j}-x_{j-1}}(y_{j}-y_{j-1})&G_{x_{j}-x_{j-1}}(y_{j}-z_{j-1})\\G_{x_{j}-x_{j-1}}(z_{j}-y_{j-1})&G_{x_{j}-x_{j-1}}(z_{j}-z_{j-1})\end{array}\bigg)\;,\nonumber
\end{align}
where
\begin{equation}
G_{x}(y)=\frac{\ee^{-\frac{y^{2}}{2x}}}{\sqrt{2\pi x}}
\end{equation}
is the Brownian propagator and the determinants come from the non-crossing condition by the method of images, see the next section.

Defining now $w(x)=\frac{w_{1}(x)+w_{2}(x)}{\sqrt{2}}$ and $e(x)=\frac{w_{1}(x)-w_{2}(x)}{\sqrt{2}}$, the change of variables $y_{j}=\frac{u_{j}+v_{j}}{\sqrt{2}}$, $z_{j}=\frac{u_{j}-v_{j}}{\sqrt{2}}$ in the expectation value of an arbitrary function of the $w_{1}(x_{j})$, $w_{2}(x_{j})$ leads for the joint probability density of $w$ and $e$ at the $x_{j}$
\begin{equation}
	q(\vec{x},\vec{u},\vec{v})=\frac{\mathbb{P}(\forall j=1,\ldots,n:\;w(x_{j})\in[u_{j},u_{j}+\dd u_{j}]\;\text{and}\;e(x_{j})\in[v_{j},v_{j}+\dd v_{j}])}{\prod_{j=1}^{n}\dd y_{j}\dd z_{j}}
\end{equation}
to a factorized expression $q(\vec{x},\vec{u},\vec{v})=p_{w}(\vec{x},\vec{u})\,p_{e}(\vec{x},\vec{v})$, indicating that the processes $w(x)$ and $e(x)$ are independent. Additionally, one finds (with the convention $u_{0}=0$)
\begin{equation}
	p_{w}(\vec{x},\vec{u})=\prod_{j=1}^{n}G_{x_{j}-x_{j-1}}(u_{j}-u_{j-1})\;,
\end{equation}
which is the joint probability density of a Wiener process, and
\begin{equation}
	p_{e}(\vec{x},\vec{v})=\frac{2v_{1}v_{n}\ee^{-\frac{v_{1}^{2}}{2x_{1}}-\frac{v_{n}^{2}}{2(1-x_{n})}}}{\sqrt{2\pi}\,x_{1}^{3/2}(1-x_{n})^{3/2}}\,\prod_{j=2}^{n}\big(G_{x_{j}-x_{j-1}}(v_{j}-v_{j-1})-G_{x_{j}-x_{j-1}}(v_{j}+v_{j-1})\big)\;,
\end{equation}
which is the joint probability density of a Brownian excursion, see e.g. \cite{DEL2004.1}.

\begin{table}
	\begin{center}
	\begin{tabular}{|l|c|}\hline&\\[-3.5mm]
		Brownian average & images\\[2.5mm]\hline&\\[-3.5mm]
		$\displaystyle\langle\max(w)\rangle=\sqrt{\frac{2}{\pi}}$ & 2\\[2.5mm]\hline&\\[-3.5mm]
		$\displaystyle\langle\max(w)^{2}\rangle=1$ & 2\\[2.5mm]\hline&\\[-3.5mm]
		$\displaystyle\langle\max(-w_{1})\max(w_{1}+w_{2})\rangle=\frac{2}{\pi}$ & 8\\[2.5mm]\hline&\\[-3.5mm]
		$\displaystyle\langle\max(-w_{1}-w_{2})\max(w_{2}+w_{3})\rangle=1$ & 6\\[2.5mm]\hline&\\[-3.5mm]
		$\displaystyle\langle\max(e-w)\rangle=\frac{2\sqrt{2}}{\sqrt{\pi}}$ & 8\\[2.5mm]\hline&\\[-3.5mm]
		$\displaystyle\langle\max(e-w)^{2}\rangle=3$ & 8\\[2.5mm]\hline&\\[-3.5mm]
		$\displaystyle\langle\max(e_{1}+e_{2}-w_{1}-w_{2})\rangle=\frac{3\sqrt{\pi}}{2}$ & 24\\[2.5mm]\hline&\\[-3.5mm]
		$\displaystyle\langle\max(e_{1}+e_{2}-w_{1}-w_{2})^{2}\rangle=8$ & 24\\[2.5mm]\hline&\\[-3.5mm]
		$\displaystyle\langle\max(e_{1}+e_{2}-w_{1}-w_{2})\max(e_{2}+e_{3}+w_{2}+w_{3})\rangle=\frac{160\pi}{27\sqrt{3}}-4$ & 720\\[2.5mm]\hline&\\[-3.5mm]
		$\displaystyle\langle\max(e_{1}+w_{1})\max(e_{1}+e_{2}-w_{1}-w_{2})\rangle\approx3.93164$ & 384\\[2.5mm]\hline
	\end{tabular}
	\end{center}
	\caption{Averages involving maxima of independent Wiener processes $w$, $w_{j}$ and Brownian excursions $e$, $e_{k}$, which are used to evaluate the first cumulants $d_{k}(0)$ for various initial conditions. The number of images needed to enforce the proper boundary conditions when solving the heat equation is given in the last column.}
	\label{table Brownian averages}
\end{table}

\subsection{Expectation values of Brownian averages}
\label{section Brownian averages}
In this section, we sketch our derivation of the Brownian averages needed to compute explicitly from (\ref{nux[h0]}) the correction $d_{k}(0)$ to the stationary cumulants for simple initial conditions $h_{0}$.

The probability density of a Wiener process $w$
\begin{equation}
	p_{w}(u)=\frac{\ee^{-\frac{u^{2}}{2x}}}{\sqrt{2\pi x}}
\end{equation}
gives
\begin{equation}
	\label{<w>}
	\langle w(x)\rangle=0
	\qquad\text{and}\qquad
	\langle w(x)^{2}\rangle=x\;.
\end{equation}
For a Brownian excursion $e$, one has instead
\begin{equation}
	p_{e}(u)=1_{\{u\geq0\}}\frac{\sqrt{2}\,u^{2}\,\ee^{-\frac{u^{2}}{2x}-\frac{u^{2}}{2(1-x)}}}{\sqrt{\pi\,x^{3}(1-x)^{3}}}\;,
\end{equation}
which leads to
\begin{equation}
	\label{<e>}
	\langle e(x)\rangle=\frac{2\sqrt{2x(1-x)}}{\sqrt{\pi}}
	\qquad\text{and}\qquad
	\langle e(x)^{2}\rangle=3x(1-x)\;.
\end{equation}

Various averages involving maxima of independent Wiener processes $w_{j}$ and Brownian excursions $e_{k}$, see table~\ref{table Brownian averages} for a summary, are additionally needed to evaluate the first cumulants $d_{k}(x)$ for various initial conditions. These Brownian averages can be computed by solving the heat equation, using the method of images to enforce the various boundary conditions required. We sketch below the method for the specific case of $\langle\max(e_{1}+e_{2}-w_{1}-w_{2})\max(e_{2}+e_{3}+w_{2}+w_{3})\rangle$. We switch to more usual variable names in the context of the heat equation.

In order to compute this average, we need the density probability
\begin{equation}
	\label{pyz1z2(t,xi)}
	p_{y,z_{1},z_{2}}(t,x_{1},\ldots,x_{6})=\frac{\mathbb{P}(w_{4},w_{5},w_{6}>y,w_{1}+w_{2}-w_{4}-w_{5}<z_{1},w_{2}+w_{3}+w_{5}+w_{6}<z_{2}),\forall i\;w_{i}(t)\in[x_{i},x_{i}+\dd x_{i}]}{\dd x_{1}\dd x_{2}\dd x_{3}\dd x_{4}\dd x_{5}\dd x_{6}}\;,
\end{equation}
with the inequalities implied for all $w_{i}(u)$, $0\leq u\leq t$. In the limit $y\to0$, where $e_{1}=w_{4}>0$, $e_{2}=w_{5}>0$, $e_{3}=w_{6}>0$, this leads to the required conditioning for the Brownian excursions, as
\begin{align}
	\label{P[max*max]}
	\mathbb{P}(e_{1}&+e_{2}-w_{1}-w_{2}<z_{1},e_{2}+e_{3}+w_{2}+w_{3}<z_{2})\\
	&=\mathbb{P}(w_{4}+w_{5}-w_{1}-w_{2}<z_{1},w_{2}+w_{3}+w_{5}+w_{6}<z_{2}|w_{4}(1)=w_{5}(1)=w_{6}(1)=0,w_{4},w_{5},w_{6}>0)\nonumber\\
	&=\lim_{\epsilon,\delta\to0^{+}}\frac{\mathbb{P}(w_{4}+w_{5}-w_{1}-w_{2}<z_{1},w_{2}+w_{3}+w_{5}+w_{6}<z_{2},w_{4}(1),w_{5}(1),w_{6}(1)\in[0,\epsilon],w_{4},w_{5},w_{6}>-\delta)}{P(w_{4}(1),w_{5}(1),w_{6}(1)\in[0,\epsilon],w_{4},w_{5},w_{6}>-\delta)}\nonumber\\
	&=\lim_{\delta\to0^{+}}\frac{\int_{-\infty}^{\infty}\dd x_{1}\dd x_{2}\dd x_{3}\,1_{\{x_{1}+x_{2}<z_{1}\}}\,1_{\{-x_{2}-x_{3}<z_{2}\}}\,p_{-\delta,z_{1},z_{2}}(1,x_{1},x_{2},x_{3},0,0,0)}{p_{-\delta}(1,0)^{3}}\;,\nonumber
\end{align}
where
\begin{equation}
	\label{py(t,x) 2 images}
	p_{y}(t,x)=\frac{\mathbb{P}(w>y,w(t)\in[x,x+\dd x])}{\dd x}=\frac{\ee^{-\frac{x^{2}}{2t}}}{\sqrt{2\pi t}}-\frac{\ee^{-\frac{(2y-x)^{2}}{2t}}}{\sqrt{2\pi t}}\propto y^{2}\;\text{when $y\to0$}\;.
\end{equation}
The function $p_{y}$ verifies the heat equation $\partial_{t}p_{y}=\frac{1}{2}\partial_{x}^{2}p_{y}$ with the initial condition $p_{y}(0,x)=1_{\{y<0\}}\,\delta(x)$ and the boundary condition $p_{y}(t,y)=0$, and the solution in (\ref{py(t,x) 2 images}) by the method of images, consisting in subtracting from the Brownian propagator $\ee^{-\frac{x^{2}}{2t}}/\sqrt{2\pi t}$ its image under the reflection $x\to2y-x$, is indeed the unique solution of the heat equation with the correct initial and boundary conditions.

The computation of the function $p_{y,z_{1},z_{2}}$ defined in (\ref{pyz1z2(t,xi)}) follows the same principle as for $p_{y}$: one has the heat equation $\partial_{t}p_{y,z_{1},z_{2}}=\frac{1}{2}(\partial_{x_{1}}^{2}+\ldots+\partial_{x_{6}}^{2})p_{y,z_{1},z_{2}}$ with the initial condition $p_{y,z_{1},z_{2}}(0,x_{1},\ldots,x_{6})=1_{\{0<y<z_{1},z_{2}\}}\,\delta(x_{1})\ldots\delta(x_{6})$ and the boundary condition $p_{y,z_{1},z_{2}}(t,x_{1},\ldots,x_{6})=0$ when either $x_{4}=y$, $x_{5}=y$, $x_{6}=y$, $x_{1}+x_{2}-x_{4}-x_{5}=z_{1}$ or $x_{2}+x_{3}+x_{5}+x_{6}=z_{2}$. From an exhaustive exploration with a computer algebra system, we observe that the group acting on $\mathbb{R}^{6}$ generated by the five corresponding reflections has 720 elements. Assigning a sign $+$ to the identity element of the group and multiplying by $-$ for each reflection eventually leads to an expression for $p_{y,z_{1},z_{2}}(t,x_{1},\ldots,x_{6})$ as a (signed) sum of reflected Brownian propagators. Taking the limit $\delta\to0$ in (\ref{P[max*max]}) finally gives, after computing the required integrals, the expression in table~\ref{table Brownian averages} for $\langle\max(e_{1}+e_{2}-w_{1}-w_{2})\max(e_{2}+e_{3}+w_{2}+w_{3})\rangle$.

\section{Bethe ansatz}
In this section, we recall known results about Bethe ansatz for TASEP with periodic and open boundaries (with $\alpha=\beta=1$), in order to explain more precisely where our conjectures (\ref{omega stat}) and (\ref{omega empty}) comes from. The derivation of our main results (\ref{nu0 stat}) and (\ref{nu0 nw}) from (\ref{omega stat}) and (\ref{omega empty}) is then detailed.

\subsection{TASEP with periodic boundaries}
For a periodic system with $L$ sites and a number $N$ of particles conserved by the dynamics, Bethe ansatz involves $N$ Bethe roots $y_{j}$, chosen as distinct solutions of
\begin{equation}
	\label{Bethe eq periodic}
	B(1-y_{j})^{L}=(-y_{j})^{N}\;.
\end{equation}
The parameter $B$ is related to the quantity conjugate to the current from a given site $i$ to $i+1$ as $B=-\ee^{\gamma}\prod_{k=1}^{N}y_{k}$, and the corresponding eigenvalue of $M_{i}(\gamma)$ is equal to $E=\sum_{j=1}^{N}\frac{y_{j}}{1-y_{j}}$.

The solutions $y_{j}(B)$ of (\ref{Bethe eq periodic}) have the branch points $B=0$, $B=B_{*}=N^{N}(L-N)^{L-N}/L^{L}$ and $B=\infty$, and for some convenient labelling of the $y_{j}$, analytic continuation across the branch cuts $(B_{*},\infty)$ and $(0,B_{*})$ is realized by the action on the index $j$ of respectively the cyclic permutation $\mathfrak{a}=(1,2,\ldots,L)$ and the product of two cyclic permutations $\mathfrak{b}=(1,\ldots,N)(N+1,\ldots,L)$ \cite{P2020.2}.

The spectral curve $\det(\lambda\Id-M_{i}(\gamma))=0$ is a polynomial equation in $\lambda$ and $g=\ee^{\gamma}$ (independent of $i$), i.e. a (complex) algebraic curve. The compact Riemann surface $\R$ associated with it can then be described as gluing together sheets $\Ch_{J}$, i.e. copies of the Riemann sphere $\Ch=\C\cup\{\infty\}$ in which the variable $B$ lives, indexed by the subsets $J$ of $\subset\{1,\ldots,L\}$ with $N$ elements specifying which Bethe roots are chosen. Analytic continuation between the sheets proceeds by applying the permutations $\mathfrak{a}$ and $\mathfrak{b}$ to the elements of $J$.

The probability of the current $Q_{0}(t)$ from site $L$ to site $1$ is then given by an integral over a contour $\Gamma\subset\R$ \cite{P2020.2}
\begin{equation}
	\label{P(Qt)}
	\P(Q_{0}(t)=Q)=\oint_{p\in\Gamma}\frac{\dd g}{2\ii\pi g}\,\frac{\mathcal{N}(p)\,\ee^{tE(p)}}{g(p)^{Q}}=\oint_{p\in\Gamma}\frac{\dd B}{2\ii\pi B}\,\exp\Big(\int_{o}^{p}t\,\dd E-Q\,\dd\gamma+\omega\Big)\;,
\end{equation}
with $\mathcal{N}=\frac{\langle\Sigma|\psi\rangle\langle\psi|\mathbb{P}_{0}\rangle}{\langle\psi|\psi\rangle}$, $\omega=\dd\log(\kappa\mathcal{N})$ and $\kappa=\frac{\dd\log g}{\dd\log B}$. The lower bound $o\in\R$ for the integral in the exponential corresponds to the stationary state at $\gamma=0$. Notably, it was observed in \cite{P2020.2} that the meromorphic differential $\omega$ has the simple expression
\begin{align}
	\label{omega stat periodic}
	\omega^{\rm stat}&=\Big(\frac{N(L-N)}{L}\,\kappa^{2}-\frac{1+g}{1-g}\,\kappa-1\Big)\frac{\dd B}{B}\\
	\label{omega dw periodic}
	\omega^{\rm dw}&=\Big(\frac{N(L-N)}{L}\,\kappa^{2}-\frac{\kappa}{1-g}\Big)\frac{\dd B}{B}\;,
\end{align}
respectively for the stationary initial condition $|\mathbb{P}_{0}\rangle=|\mathbb{P}_{\rm st}\rangle$ (where all the configurations have the same probability for periodic TASEP), and the (deterministic) domain wall state with sites $L-N+1$ through $L$ occupied.

\subsection{Open TASEP with $\alpha=\beta=1$}

\subsubsection{Analytic continuations in the variable $B$}


Inserting $w=(y+y^{-1})/2$ into (\ref{Bethe eq w}), the $y_{j}^{\pm}$ are solution of the algebraic equation $P(y_{j}^{\pm},B)=0$ with
\begin{equation}
	\label{Bethe eq y}
	P(y,B)=B\,(1-y)^{2L+2}(1+y)^{2}+(-y)^{L+2}\;.
\end{equation}
Solving the system $P(y,B)=0$, $\partial_{y}P(y,B)=0$, it follows that the $y_{j}^{\pm}(B)$ have the branch points $0$, $B_{*}$, $\infty$ with
\begin{equation}
	B_{*}=\frac{(L+2)^{L+2}}{2^{2L+4}(L+1)^{L+1}}\;.
\end{equation}



Similarly to TASEP with periodic boundaries, the Riemann surface $\R$ associated with the spectral curve $\det(\lambda\Id-M_{i}(\gamma))=0$ for open TASEP, independent of $i$ since $M_{i}(\gamma)$ and $M_{0}(\gamma)$ are related by a similarity transformation, is built from sheets $\Ch_{\Theta}$ for all $\Theta=(\theta_{0},\ldots,\theta_{L+1})$ verifying the condition $\prod_{j=0}^{L+1}\theta_{j}=(-1)^{L}$. The $2^{L+1}$ sheets are glued together in $\R$ according the action of the permutations $\mathfrak{a}$ and $\mathfrak{b}$ on $\Theta$.

\subsubsection{Differential $\omega$}
The expression (\ref{P(Qt)}) for the probability of the current still holds in principle for open TASEP. It is then natural to seek for $\omega=\dd\log(\kappa\,\mathcal{N})$ with $\mathcal{N}=\frac{\langle\Sigma|\psi\rangle\langle\psi|\mathbb{P}_{0}\rangle}{\langle\psi|\psi\rangle}$ simple expressions similar to (\ref{omega stat periodic}) for simple initial conditions. Setting again $g=\ee^{\gamma}$, extensive numerics indicate that
\begin{align}
	\label{omega stat g}
	\omega^{\rm stat}&=\Big((L+2)\,\kappa^{2}-\frac{1+g}{1-g}\,\kappa-1\Big)\frac{\dd B}{B}\\
	\label{omega empty g}
	\omega^{\rm empty}&=\Big((L+2)\,\kappa^{2}-\frac{\kappa}{1-g}-\kappa\Big)\frac{\dd B}{B}\\
	\label{omega full g}
	\omega^{\rm full}&=\Big((L+2)\,\kappa^{2}-\frac{\kappa}{1-g}-\frac{L+2}{2}\,\kappa\Big)\frac{\dd B}{B}\;,
\end{align}
hold respectively for stationary initial condition, and deterministic initial conditions with a system either totally empty of filled with particles. For stationary initial condition, the only difference with the corresponding result (\ref{omega stat periodic}) for periodic boundaries (except for the fact that the Bethe equations differ, and hence also $\kappa=\frac{\dd\log g}{\dd\log B}$) is that the coefficient $N(L-N)/L$ is replaced by $L+2$. The two other expressions resemble the one in (\ref{omega dw periodic}).

\subsubsection{Computation of $\kappa_{\rm st}$ at large $L$}
From (\ref{Bethe eq y}), the function $\kappa=\frac{\dd\log g}{\dd\log B}$ can be written in terms of the Bethe roots as
\begin{equation}
	\label{kappa[y]}
	\kappa=\sum_{j=0}^{L+1}\frac{1}{(L+2)(1+y_{j}^{-1})-4(1+y_{j})^{-1}}\;.
\end{equation}
The stationary eigenstate corresponds to choosing the Bethe roots $y_{j}=y_{j}^{-}=w_{j}-w_{j}\sqrt{1-w_{j}^{-2}}$, which vanish when $B\to0$ (corresponding to $w_{j}\to\infty$), unlike the $y_{j}^{+}$. Then, following \cite{DL1998.1,CN2018.1}, one can write $\kappa$ for the stationary eigenstate at small enough $|B|$ as the contour integral around $0$ with positive orientation
\begin{equation}
	\kappa_{\rm st}=\oint\frac{\dd y}{2\ii\pi}\frac{1}{(L+2)(1+y^{-1})-4(1+y)^{-1}}\,\partial_{y}\log\Big(B\,(1-y)^{2L+2}(1+y)^{2}+(-y)^{L+2}\Big)\;,
\end{equation}
where we used (\ref{kappa[y]}) and (\ref{Bethe eq y}). Expanding in powers of $B$ and evaluating the contour integral by residues leads after some calculations to
\begin{equation}
	\kappa_{\rm st}=-\frac{1}{2}\sum_{n=1}^{\infty}\frac{{{2(L+1)n}\choose{Ln}}{{2n}\choose{n}}}{{{(L+1)n}\choose{n}}}\,B^{n}\;,
\end{equation}
which has for finite $L$ a radius of convergence equal to $B_{*}$. Setting $B=-B_{*}\,\ee^{v-1}$, Stirling's formula eventually leads at large $L$ to
\begin{equation}
	\label{kappa[chi'']}
	\kappa_{\rm st}\simeq\frac{2\chi''(v)}{\sqrt{L}}
\end{equation}
with $\chi$ as in (\ref{chi SM}) and $v<1$, as stated in the letter.

\subsubsection{Computation of $\mathcal{N}_{\rm st}$ at large $L$}
\label{section Nst asymptotics}
We consider again the function
\begin{equation}
	\mathcal{N}_{\rm st}(\gamma)=\frac{\langle\Sigma|\psi_{\rm st}(\gamma)\rangle\langle\psi_{\rm st}(\gamma)|\mathbb{P}_{0}\rangle}{\langle\psi_{\rm st}(\gamma)|\psi_{\rm st}(\gamma)\rangle}\;.
\end{equation}
Since $|\psi_{\rm st}(\gamma)\rangle\propto|\mathbb{P}_{0}\rangle$ and $\langle\psi_{\rm st}(\gamma)|\propto\langle\Sigma|$, we observe that
\begin{equation}
	\label{N(0)}
	\mathcal{N}_{\rm st}(0)=1\;.
\end{equation}

We would like to compute the large $L$ asymptotics of $\mathcal{N}_{\rm st}(\gamma)$ from the conjectured exact expressions (\ref{omega stat g})-(\ref{omega full g}) for special initial conditions. We set $B=-B_{*}\,\ee^{v-1}$ and scale $\gamma=\log g$ as in (\ref{E[mu] ; gamma[s]}), with $s$ and $v$ related through $s=\chi'(v)$. Then, (\ref{kappa[chi'']}) combined with (\ref{omega stat g})-(\ref{omega full g}) leads at large $L$ for the stationary eigenstate to
\begin{align}
	\omega_{\rm st}^{\rm stat}&\simeq\Big(4\chi''(v)^{2}+\frac{2\chi''(v)}{\chi'(v)}-1\Big)\,\dd v\\
	\omega_{\rm st}^{\rm empty}\simeq\omega_{\rm st}^{\rm full}+\sqrt{L}\,\chi''(v)\,\dd v&\simeq\Big(4\chi''(v)^{2}+\frac{\chi''(v)}{\chi'(v)}\Big)\,\dd v\;.
\end{align}
Integrating $\omega_{\rm st}=\dd\log(\kappa_{\rm st}\,\mathcal{N}_{\rm st})$ from $\gamma_{1}$ to $\gamma_{2}$, corresponding through (\ref{E[mu] ; gamma[s]}) to $s_{1}=\chi'(v_{1})$, $s_{2}=\chi'(v_{2})$, one has
\begin{align}
	\frac{\kappa_{\rm st}(\gamma_{2})\,\mathcal{N}_{\rm st}^{\rm stat}(\gamma_{2})}{\kappa_{\rm st}(\gamma_{1})\,\mathcal{N}_{\rm st}^{\rm stat}(\gamma_{1})}&\simeq\ee^{v_{1}-v_{2}}\,\frac{\chi'(v_{2})^{2}}{\chi'(v_{1})^{2}}\,\exp\Big(4\int_{v_{1}}^{v_{2}}\dd v\,\chi''(v)^{2}\Big)\\
	\frac{\kappa_{\rm st}(\gamma_{2})\,\mathcal{N}_{\rm st}^{\rm empty}(\gamma_{2})}{\kappa_{\rm st}(\gamma_{1})\,\mathcal{N}_{\rm st}^{\rm empty}(\gamma_{1})}\simeq\ee^{\sqrt{L}\,(s_{2}-s_{1})}\,\frac{\kappa_{\rm st}(v_{2})\,\mathcal{N}_{\rm st}^{\rm full}(v_{2})}{\kappa_{\rm st}(v_{1})\,\mathcal{N}_{\rm st}^{\rm full}(v_{1})}&\simeq\frac{\chi'(v_{2})}{\chi'(v_{1})}\,\exp\Big(4\int_{v_{1}}^{v_{2}}\dd v\,\chi''(v)^{2}\Big)\;.
\end{align}
We use again (\ref{kappa[chi'']}) to express $\mathcal{N}_{\rm st}$ alone in terms of $\chi$ and its derivatives, as
\begin{align}
	\mathcal{N}_{\rm st}^{\rm stat}(\gamma_{2})&\simeq\frac{\ee^{v_{1}}\,\chi''(v_{1})\,\mathcal{N}_{\rm st}^{\rm stat}(\gamma_{1})}{\chi'(v_{1})^{2}}\times\frac{1}{\ee^{v_{2}}}\,\frac{\chi'(v_{2})^{2}}{\chi''(v_{2})}\,\exp\Big(4\int_{v_{1}}^{v_{2}}\dd v\,\chi''(v)^{2}\Big)\\
	\mathcal{N}_{\rm st}^{\rm empty}(\gamma_{2})\simeq\ee^{\sqrt{L}\,(s_{2}-s_{1})}\,\mathcal{N}_{\rm st}^{\rm full}(\gamma_{2})&\simeq\frac{\chi''(v_{1})\,\mathcal{N}_{\rm st}^{\rm stat}(\gamma_{1})}{\chi'(v_{1})}\times\frac{\chi'(v_{2})}{\chi''(v_{2})}\,\exp\Big(4\int_{v_{1}}^{v_{2}}\dd v\,\chi''(v)^{2}\Big)\;.
\end{align}
We now take the limit $\gamma_{1}\to0$, corresponding to $s_{1}\to0$ and $v_{1}\to-\infty$. Using (\ref{N(0)}) and $\chi'(v_{1})\simeq\chi''(v_{1})\simeq\frac{\ee^{v}}{8\sqrt{\pi}}$, we finally obtain
\begin{align}
	\label{Nst stat}
	\mathcal{N}_{\rm st}^{\rm stat}(\gamma_{2})&\simeq\frac{8\sqrt{\pi}}{\ee^{v_{2}}}\,\frac{\chi'(v_{2})^{2}}{\chi''(v_{2})}\,\exp\Big(4\int_{-\infty}^{v_{2}}\dd v\,\chi''(v)^{2}\Big)\\
	\label{Nst empty full}
	\mathcal{N}_{\rm st}^{\rm empty}(\gamma_{2})\simeq\ee^{\sqrt{L}\,s_{2}}\,\mathcal{N}_{\rm st}^{\rm full}(\gamma_{2})&\simeq\frac{\chi'(v_{2})}{\chi''(v_{2})}\,\exp\Big(4\int_{-\infty}^{v_{2}}\dd v\,\chi''(v)^{2}\Big)\;.
\end{align}
The extra factor $\ee^{\sqrt{L}\,s_{2}}=\ee^{-\gamma_{2}\,D\,L}$, where $D=-1/2$ is the constant appearing in (\ref{H[h,D]}) for the initial condition where the system is full, cancels when computing the generating function (\ref{GF[h]}).

\begin{thebibliography}{10}
\bibitem{SM}
	See {S}upplementary {M}aterial, which includes {R}efs. [34,35], and contains a
	more detailed derivation of the results presented in the letter.
\end{thebibliography}
\end{document}